\documentclass[prodmode,acmtecs]{acmsmall} 

\usepackage[ruled]{algorithm2e}
\renewcommand{\algorithmcfname}{ALGORITHM}
\SetAlFnt{\small}
\SetAlCapFnt{\small}
\SetAlCapNameFnt{\small}
\SetAlCapHSkip{0pt}
\IncMargin{-\parindent}

\usepackage{amsmath,amssymb}
\usepackage{fancyhdr}
\usepackage{fancybox}
\usepackage{shadow}
\usepackage{setspace}
\usepackage{epsfig}
\usepackage{subfigure}
\usepackage{epstopdf}
\usepackage{graphicx}
\usepackage{graphics}

\usepackage{amsmath,amssymb}
\usepackage{fancyhdr}
\usepackage{fancybox}
\usepackage{shadow}
\usepackage{setspace}
\usepackage{epsfig}
\usepackage{subfigure}
\usepackage{epstopdf}
\usepackage{graphicx}
\usepackage{graphics}

\newtheorem{ex}{Example}

\begin{document}

\markboth{Hassaan, M. and Gouda, K.}{New Subgraph Isomorphism Algorithms: Vertex versus Path-at-a-time Matching}

\title{New Subgraph Isomorphism Algorithms: Vertex versus
Path-at-a-time Matching}
\author{  Mosab Hassaan
\affil{Faculty of Science, Benha University, Egypt
}
Karam Gouda
\affil{Faculty of Computers \& Informatics, Benha University, Egypt }
}

\begin{abstract}
Graphs are widely used to model complicated data semantics in many application domains. In this paper,
two novel and efficient algorithms {\tt Fast-ON} and {\tt Fast-P} are proposed for solving the subgraph isomorphism
problem. The two algorithms are based on Ullman algorithm \cite{ullman}, apply vertex-at-a-time match-
ing manner and path-at-a-time matching manner respectively, and use effective heuristics to cut the
search space. Comparing to the well-known algorithms, {\tt Fast-ON} and {\tt Fast-P} achieve up to 1-4 orders of
magnitude speed-up for both dense and sparse graph data.

\end{abstract}








\maketitle

\section{Introduction}
As a popular data structure, graphs have been used to model many
complex data objects and their relationships in the real world, such
as the chemical compounds \cite{chem}, entities in images
\cite{image}, and social networks \cite{social}. For example, in
social network, a person $i$ corresponds to a vertex $v_i$ in the
graph $G$, and another person $j$ corresponds to a vertex $v_j$ in
the graph $G$. If persons $i$ and $j$ are acquaintances or they have
a business relation, then an edge $(v_i, v_j)$ exists, which
connects vertex $v_i$ and $v_j$. Also in chemistry, a set of atoms
combined with designated bonds are used to describe chemical
molecules.\\

{\em Subgraph isomorphism} is an important and very general form of
pattern matching that finds practical applications in areas such as
pattern recognition and computer vision, computer-aided design,
image processing, graph grammars, graph transformation, bio
computing, search operation in chemical structural formulae
database, and numerous others. Moreover, subgraph isomorphism
checking is the basic and important operation in managing and
analyzing graph data. In other words, it is the building block of
many graph analysis and management activities. For example, in {\em
Frequent Subgraph Mining} -- a well-addressed problem in graph data
analysis -- the objective  is to extract all subgraphs in a given set
of data graphs, that occur in at least a specified number of data
graphs. The core in solving this problem is subgraph isomorphism
checking. The reason is given as follows. One main challenge
in frequent subgraph mining is to count how many data graphs
containing each given candidate subgraph. This involves subgraph
isomorphism checking  between  the candidates and each data graph.
Another example is the well-known {\em Subgraph Search}, an
important problem  in graph data management. The objective of subgraph search is to retrieve data graphs that
contain a query graph as a subgraph. Subgraph isomorphism checking
plays an important role in any solution to this problem.\\


Informally, two graphs $H$ and $G$ are isomorphic if it is
possible to redraw one of them, say $G$, so it appears to be
identical to $H$. In other words, it asks whether there is a
one-to-one mapping between the vertices of the two graphs,
preserving vertex connections (the edges). On the other hand, the
subgraph isomorphism problem asks the following question. Given two
graphs $H$ and $G$, is $H$ isomorphic to any subgraph of $G$?
Graph isomorphism is neither known to be solvable in polynomial time nor
NP-complete, while subgraph isomorphism is
known to be NP-complete \cite{np}.\\

{\bf{\emph{Contribution.}}}
In this paper, we propose two new algorithms for subgraph
isomorphism checking. These algorithms are based on Ullman algorithm
and improve upon it by reducing its search space.  The first
algorithm reduces the search space size by utilizing the label
information of vertex's neighborhood, and speeding up the search by
following a novel ordering strategy of the query's vertices. The
algorithm is called {\tt Fast-ON}.
Comparing to the well-known algorithms Ullman \cite{ullman} and Vflib \cite{vflib},
{\tt Fast-ON} achieves up to 1-3 orders of magnitude speed-up.\\

The second algorithm explores the possibility of leveraging
substructural  matching instead of vertex matching. 
In fact, substructure matching will cut down the depth of the search
tree, and reduce the search size as  the matching candidates
will also be minimized accordingly. This new algorithm follows a path-at-a-time
matching manner, and called {\tt Fast-P}.
To speed up the search in {\tt Fast-P}, we propose an ordering of
the query paths to force false mappings to be discarded as early as
possible during the search. Comparing to the well-known algorithms
Ullman \cite{ullman}  and Vflib \cite{vflib}, {\tt Fast-P} achieves
up to 1-4 orders of magnitude speed-up.\\

{\bf{\emph{Organization.}}} This paper is organized as follows. Section \ref{sec:preconcepts} defines the preliminary and concepts. Section \ref{sec:relatedwork} presents the related work. Section \ref{sec:fast-on-p1} presents our two new algorithms ({\tt Fast-ON} and {\tt Fast-P}). Section \ref{sec:experiments} reports the experimental results. Finally, Section \ref{sec:Con} concludes the paper.

\section{Preliminaries}
\label{sec:preconcepts}
In this section, we introduce the fundamental concepts. Let $\Sigma$ be a set of discrete-valued labels. A labeled graph is
a 3-tuple, $G = (V_G, E_G, l_G)$ where  $V_G$ is a set of vertices. Each $v \in V_G$ is a unique ID
representing a vertex, $E_G \subseteq V_G \times V_G$ is a set of edges (directed or undirected), and
$l_G: V_G \cup E_G \longmapsto \Sigma$ is a function assigning labels to the vertices and edges of the graph. A labeled graph $G$ is said to be \textit{connected},
if  each pair of vertices  $v_i, v_j$ $\in$  $V_G$, $i\neq j$, are directly or indirectly connected.
This paper focuses on undirected, simple (no self-loops, no duplicate edges), labeled, and connected graphs. Given a
graph $G$, we define the set of adjacent vertices (or neighbors) of a vertex $v \in G$ as
$adj_G (v) = \{u : (v, u) \in E_{G}\}$, and the degree of $v$ as $deg_G(v) = |adj_G(v)|$. The size of $G$ is
denoted by $|G| = |E_G|$. In what follows,  a labeled graph is simply called a graph unless stated otherwise.

\begin{definition} {\bf Labeled Paths.}
A path $p = u \rightsquigarrow  u^\prime$ from a vertex $u$ to a
vertex $u^\prime$ in a labeled graph $G$ is a sequence $v_0, v_1, \ldots,
v_k$ of vertices such that $u = v_0$ and $u^\prime = v_k$, and
$(v_{i-1}, v_i) \in E_G$ $\forall i= 1\ldots k$. In other words,
it is a sequence of edges connecting two vertices $u \in V_G$, $u^\prime
\in V_G$. If the vertex label is used instead of its id, for each vertex in
the path, the path is called {\em labeled} path.
$\blacksquare$
\end{definition}

A path without repetitive vertices is often  referred to as a {\em
simple} path. A {\em cycle} is a special path with at least three
edges, in which the first and last vertices are identical, but
otherwise all vertices are distinct.
\begin{definition} {\bf Graph Isomorphism.}
Given two graphs $H = (V_H, E_H, l_H)$ and $G = (V_G, E_G, l_G)$.
A graph isomorphism from $H$ to $G$ is a bijection $f: V_H \longmapsto V_G$ such that:
\begin{enumerate}
\item $(u, v) \in E_H$ iff $(f(u), f(v)) \in E_G$,
\item $l_{H}(u) = l_{G}(f(u))$ $\forall u \in V_H$, and
\item $l_{H}((u, v))=l_{G}((f(u), f(v)))$. $\blacksquare$
\end{enumerate}
\end{definition}
\vspace{.5cm}

In other words, the isomorphism $f$ preserves the edge adjacencies,
as well as the vertex and edge labels. If the function $f$ is only
injective but not bijective, we say that $H$ is isomorphic to a
subgraph of $G$, or subgraph isomorphic to G, denoted $H \subseteq
G$. In this case we also say that $G$ contains $H$.\\

A {\em graph automorphism} is an isomorphism from the graph to
itself. Given a graph $G$, the group of all its isomorphic graphs
are called an automorphism group. The graph $G$ may also contain
many occurrences (embeddings) of the subgraph $H$. Two embeddings
are considered {\em redundant} if their corresponding subgraphs are
automorphic.
\begin{figure}[h]
  \centerline{\psfig{figure=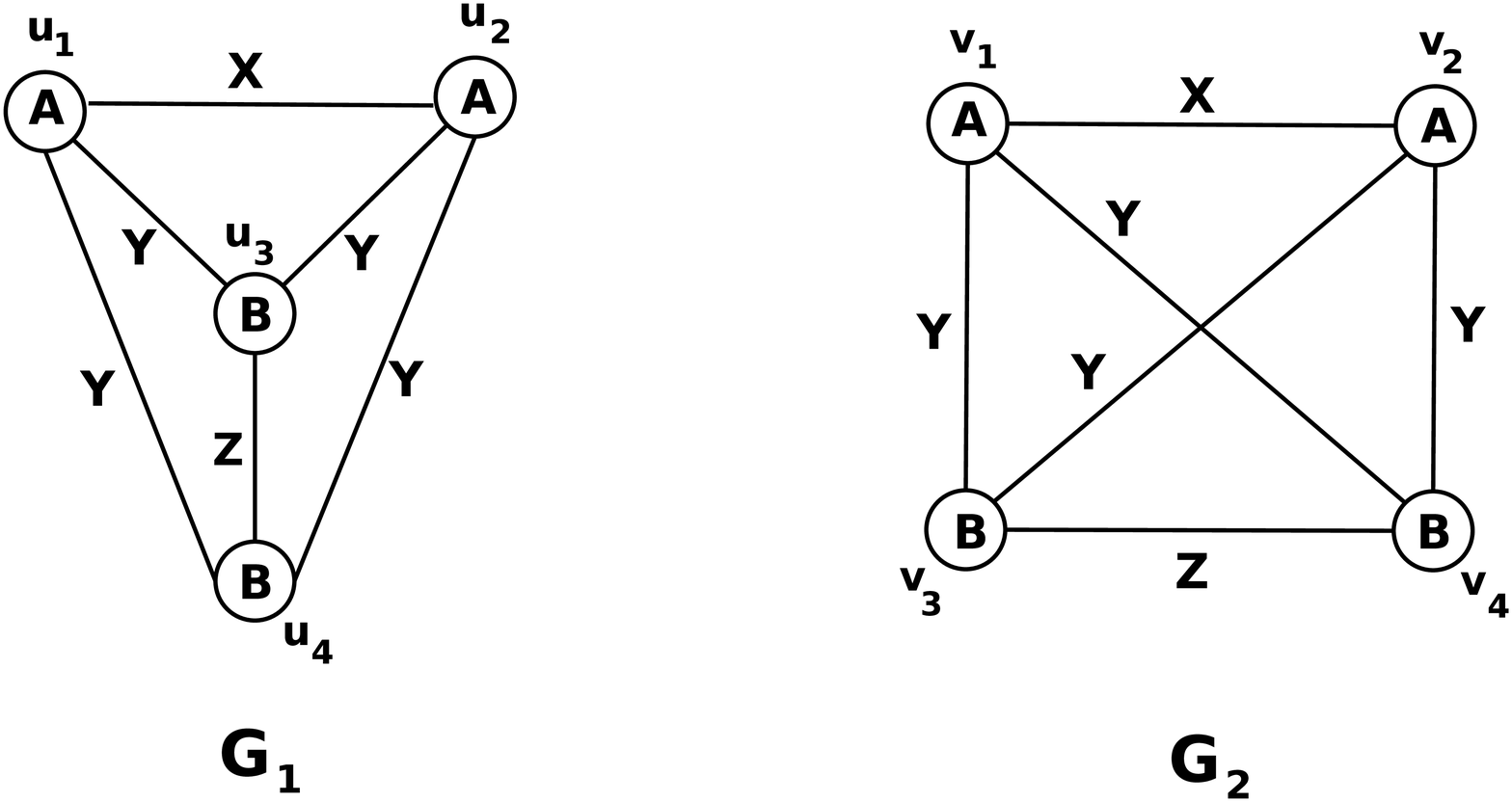,width=2.9in,height=1.5in} }
  \caption{Graph Isomorphism}
  \label{fig:graphISO}
\end{figure}
\begin{figure}[h]
  \centerline{\psfig{figure=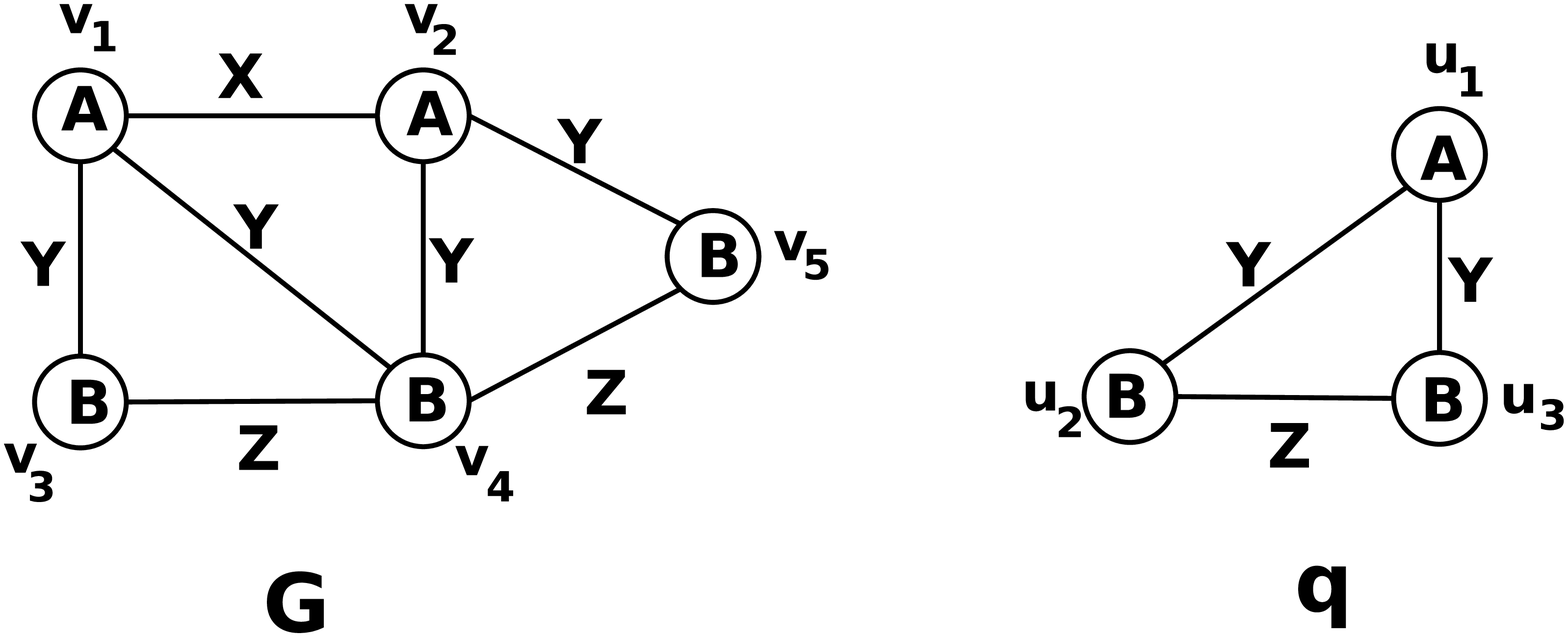,width=2.5in,height=1.2in} }
  \caption{Subgraph Isomorphism}
  \label{fig:graphSUBISO}
\end{figure}
\begin{figure}[h]
\centering
\subtable[]{
    \begin{tabular}{|c|c|c|c|c|}
    \hline
                          & $u_1$             & $u_2$        & $u_3$   & $u_4$  \\\hline
    $f_1$            & $v_1$             & $v_2$        & $v_4$   & $v_3$  \\
    $f_2$    & $v_2$             & $v_1$        & $v_3$   & $v_4$  \\\hline
   \end{tabular}
} \hspace{2cm} \subtable[]{
    \begin{tabular}{|c|c|c|c|}
    \hline
                          & $u_1$             & $u_2$        & $u_3$            \\\hline
    $f_1$            & $v_1$             & $v_3$        & $v_4$            \\
    $f_2$    & $v_1$             & $v_4$        & $v_3$   \\
        $f_3$            & $v_2$             & $v_4$        & $v_5$            \\
            $f_4$            & $v_2$             & $v_5$        & $v_4$                  \\\hline
   \end{tabular}
}
    \caption{The Set of all Possible Graph (a) and Subgraph Isomorphism (b)}
       \label{fig:Graph(SUB)ISO}
\end{figure}
\begin{ex}
In Figure \ref{fig:graphISO}, $G_1$ and $G_2$ are isomorphic graphs.   An
example of an isomorphism is $f(v_1) = u_1$, $f(v_2) = u_2$, $f(v_3) =
u_3$, and $f(v_4) = u_4$.  In Figure \ref{fig:graphSUBISO}, $q$ is
subgraph isomorphic to $G$.  An example of an subgraph isomorphism is
$f(u_1) = v_1$, $f(u_2) = v_3$ and $f(u_3) = v_4$. There are several
possible graph or subgraph isomorphisms between two graphs. The set
of all possible graph isomorphisms from $G_1$ to $G_2$ are shown in
Figure \ref{fig:Graph(SUB)ISO}(a). Also, the set of all possible
subgraph isomorphisms from $q$ to $G$ are shown in Figure
\ref{fig:Graph(SUB)ISO}(b). The subgraphs identified by the two
mappings $f_1$ and $f_2$ are redundant. So $f_3$ and $f_4$. $\blacksquare$
\end{ex}

\section{Related Work}
\label{sec:relatedwork}
A straightforward approach to check subgraph isomorphism between
the graph query $q$ against a data graph $G$ is to explore a
tree-structured search space considering all possible
vertex-to-vertex correspondences from $q$ to $G$. The search space
traversal is halted until the structure of $q$ implied by the vertex
mapping does not correspond in $G$, while reaching a leaf node of
the search space means successfully mapping all vertices of $q$ upon
$G$ without violating the structure and label constraints of
subgraph isomorphism, and it is, therefore, equivalent to having found a
matching of $q$ in $G$.\\

\begin{figure}[h]
\centering
\epsfig{file=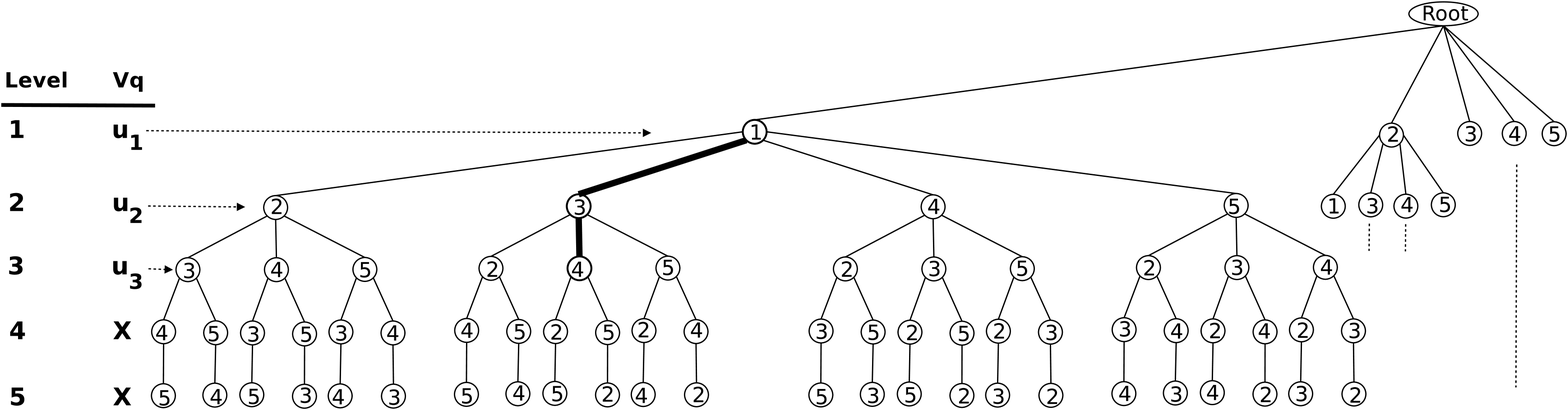,width=5.5in,height=2.7in}
\caption{A Part of Search Tree} \label{fig:searchtree1}
\end{figure}

The tree in Fig. \ref{fig:searchtree1} shows a part of the search
space generated from testing the two graphs $q$ and $G$ in Fig.
\ref{fig:graphSUBISO} for subgraph isomorphism. This space
enumerates  all possible mappings between the vertices of the two
graphs.  At level $i$ of the tree, a vertex $u_i$ in $V_q$ is mapped
to some vertex in $V_G$ (the number $j$ inside each node in the
search tree means that this node represents the vertex $v_j \in
V_G$). The root node of the search tree represents the starting
point of the search, inner nodes of the search tree correspond to
partial mappings, and nodes at level $|V_q|$ represent complete --
not necessarily sub-isomorphic -- mappings. If there exists a
complete mapping that preserves adjacency in both $q$ and $G$, then
we have $q$ is subgraph isomorphic to $G$, otherwise $q$ is not
subgraph isomorphic to $G$. The bold path in the tree, ($u_1$ is
mapped to $v_1$, $u_2$ is mapped to $v_3$, and $u_3$ is mapped to
$v_4$), is a complete mapping that preserves adjacency in $q$ and
$G$, thus $q$ is subgraphs isomorphic
to $G$.\\

\begin{definition} {\bf Matching Candidate Set.}
Given a vertex $u \in V_q$, the matching  candidate of $u$ is a set
$Cand(u)$ of vertices in $G$ sharing the same vertex label with $u$,
i.e., $Cand(u)$ = $\{v \in V_G : l_q(u) = l_G(v) \}$. $\blacksquare$
\end{definition}

Thus, in the naive approach, for each vertex $u \in V_q$, an
exhaustive search of possible one-to-one correspondences to $v \in
Cand(u)$ is required. Therefore, the total search space of the naive
algorithm is $\prod_{i=1}^{N} Cand(u_i)$, where $N = |V_q|$.  The
worst-case time complexity of the algorithm is $O(M^N)$, where $M =
|V_G|$ and $N = |V_q|$. This is a consequence of subgraph
isomorphism that is known to be NP-complete. In practice, the
running time  depends tightly on the size of the search space,
$\prod_{i=1}^{N} |Cand(u_i)|$.

\subsection{Ullman Algorithm}
Ullman algorithm \cite{ullman} is the earliest and highly-cited approach to the
subgraph isomorphism problem. Given a query graph $q$ and a data
graph $G$.  To check if $q$ is subgraph of $G$, Ullman's basic
approach is to enumerate all possible mappings of vertices in $V_q$
to those in $V_G$ using a depth-first tree-search algorithm. In
order to cope with subgraph isomorphism problem efficiently, Ullman
proposed a {\em refinement procedure} to prune the search space. It is
based on the following three conditions:
\begin{enumerate}
\item {\bf Label and degree condition.} A vertex $u \in V_q$ can be
mapped to $v \in V_G$ under injective mapping $f$, i.e $v = f(u)$,
if\\ (i) $l_{q}(u)$ =  $l_{G}(v)$, and\\ (ii) $deg_q(u) \leq
deg_G(v)$.

\item {\bf One-to-One mapping of vertices condition.} Once a vertex $u \in V_q$
is mapped to $v \in V_G$, we cannot map any other vertex in $V_q$ to
the vertex $v$.

\item {\bf Neighbor condition.} By this condition Ullman  algorithm
examines the feasibility of mapping $u \in V_q$ to $v \in V_G$ by
considering the preservation of structural connectivity. If there
exist edges connecting $u$ with previously explored vertices of $q$
but there are no  counterpart edges in $G$, the mapping test simply
fails.
\end{enumerate}
Applying the above three conditions, $|Cand(u)|$ for each $u \in
V_q$ could be decreased; thus cutting down the search space.\\



\subsection{QuickSI Algorithm}
\label{subsec:quicksi-alg} QuickSI \cite{Tamming} is a recent
subgraph checking algorithm. It is based on Ullman, and improve upon
it by speeding up Ullman's search. The underlying observation behind
developing QuickSI algorithm is noting that the Ullman's search is
random. Ullman usually matches query vertices in the input order. Some orderings do not preserve
connectivity between consecutive query vertices, which requires
Ullman to consume a lot of time checking the feasibility of partial
mappings.  Instead of trivially enumerating mappings according to
the given order of $V_q$, QuickSI enumerates mappings from a
spanning tree of $V_q$ to $V_G$ to reduce the combinations by the
connectivity restriction.\\

QuickSI proposes to follow a search order given by the
$QI$-$Sequence$. $QI$-$Sequence$ is a sequence that represents a
rooted spanning tree, $t_q$, for $q$ and  consists of a list of
spanning entries, $T_i$ , for $1 \leq i \leq |V_q|$, where each
$T_i$ keeps the basic information of the spanning tree of $q$. In
$QI$-$Sequence$, a $T_i$ may be followed by a list of extra entries,
$R_{ij}$, which keeps the extra topology information related to the
corresponding spanning
entry.
To identify a subgraph isomorphic mapping from $q$ to $G$, QuickSI
iteratively grows each possible mapping on $t_q$ in a depth-first
manner according to the vertices order in $QI$-$Sequence$. QuickSI
can terminate earlier if a prefix of $QI$-$Sequence$ cannot be
sub-isomorphically mapped to $G$. To effectively reduce the search
costs, the authors propose to reorder the $QI$-$Sequence$ as
follows. Pick up the vertex $v$ from $q$, such that its label has
the lowest occurrence in the graph $G$, as the the first entry in
$QI$-$Sequence$.  Then, iteratively pick up an unchosen vertex such
that the spanning edge has the lowest occurrence in the graph $G$
among all valid options.

\subsection{Vflib Algorithm}
The Vflib algorithm \cite{vflib} is another important algorithm for
subgraph isomorphism problem. It uses a different strategy  from
Ullman algorithm. Vflib proceeds by creating and modifying a match
state. The match state contains a matched-set, which is a set of
vertex pairs that match between the query graph $q$ and  data graph
$G$. If the matched-set contains all of the query graph $q$, then
the algorithm is successful and returns. Otherwise, the algorithm
attempts to add a new pair. It does this by tracking  the set of
vertices immediately adjacent to the matched-set. This set defines
the potential vertices that can be added to a given state. The only
pairs that can be added are either in the adjacent sets of both
graphs. The algorithm uses backtracking search to find either a
successful match state, or return a failure.

\section{New Subgraph Isomorphism  Algorithms}
\label{sec:fast-on-p1}
Clearly, the subgraph isomorphism checking is very costly, and it
becomes even challenging when the graph and the query are large and
dense. In order to alleviate the time consuming search considered by
previous algorithms, we consider reducing the search space size
$\prod_{i=1}^{N} |Cand(u_i)|$ in the following two aspects:

\begin{itemize}
\item Minimize $|Cand(u)|$ for each vertex $u\in V_q$.
\item Minimize the number of one-to-one correspondence checking,
i.e., minimize $N$.
\end{itemize}

In this paper, we propose two new algorithms for subgraph
isomorphism checking. These algorithms are based on Ullman algorithm
and improve upon it by reducing its search space.  The first
algorithm reduces the search space size by utilizing the label
information of vertex's neighborhood, and speeding up the search by
following a novel ordering strategy of the query's vertices. The
algorithm is called {\tt Fast-ON} (which stands for the bold letters
in: {\tt Fast} subgraph testing by {\tt O}rdering the query's
vertices and utilizing labeled {\tt N}eighborhood information).
Comparing to the well-known algorithms Ullman \cite{ullman} and Vflib \cite{vflib},
{\tt Fast-ON} achieves up to 1-3 orders of magnitude speed-up. {\tt
Fast-ON} algorithm is published in \cite{karam-mosab1}.\\

The second algorithm explores the possibility of leveraging
substructural  matching instead of vertex matching to minimize $N$.
In fact, substructure matching will cut down the depth of the search
tree, and consequently the search size as  the matching candidates
will also be minimized. This new algorithm follows a path-at-a-time
matching manner, and called {\tt Fast-P} which stands for the bold
letters in: {\tt Fast P}ath-at-a-time manner. To speed up the search
in {\tt Fast-P}, we propose an ordering of the query paths to force
false mappings to be discarded as early as possible during the
search. In Section \ref{sec:substructure}, {\tt Fast-P} algorithm is
discussed in details. Next, we introduce {\tt Fast-ON} algorithm.

\subsection{{\tt Fast-ON} Algorithm }
\label{subsec:fast-on}

The search space considered by Ullman algorithm is still huge even
after using the refinement procedure. {\tt Fast-ON} explores much
smaller space than that of Ullman algorithm by utilizing vertex neighborhood as in the following optimization.

\subsubsection{Opt1: Utilizing Neighborhood Labels}
\label{subsec:NeighborhoodLabels}

Here, we introduce a condition effective in reducing the
search space. It is based on the neighborhood labels of matching
vertices. This new condition is much stronger than the label and
degree condition of the refinement procedure in  Ullman algorithm.
First, we define the labeled neighborhood of any vertex as follows.

\begin{definition} {\bf Vertex Labeled Neighborhood.}
Given a graph $G$ and a vertex $u\in V_G$,  the labeled neighborhood
of $u$  is given as $NL_G(u)$ = $\{(l_{G}(v), l_{G}((u,v))): v \in
V_G$ and $(u, v) \in E_G\}$.  $\blacksquare$
\end{definition}

The following theorem presents the necessary  condition required to
map a vertex $u\in V_q$ to a vertex $v\in V_G$.
\begin{theorem}
\label{thm:containmentneighbor} Given two graphs $q$ and $G$ such
that  $q$ is subgraph isomorphic $G$ under injective function f. If
$u \in V_q$ is mapped to $v \in V_G$, then $NL_q(u) \subseteq
NL_G(v)$.   $\blacksquare$
\end{theorem}

Thus, according to Theorem~\ref{thm:containmentneighbor},  if the
labeled neighborhood of a vertex $v\in V_G$ does not contain the
labeled neighborhood of a vertex $u\in V_q$, $u$ can not be mapped to
$v$.  We can reduce the search space by enforcing this inclusion
test. Next condition generalizes the first condition of the
refinement procedure in  Ullman algorithm by adding this
inclusion test.

\begin{enumerate}
\item {\bf Label and neighborhood inclusion condition.}
A vertex $u \in V_q$ can be mapped to $v \in V_G$
under injective function $f$, i.e $v = f(u)$,
if\\
(i) $l_{q}(u)$ = $l_{G}(v)$, and\\
(ii) $NL_q(u) \subseteq NL_G(v)$.
\end{enumerate}

Note that if $NL_q(u) \subseteq NL_G(v)$ is  satisfied, it directly
leads to $deg(u) \leq deg(v)$ since $deg(v) = |NL_G(v)|$ for simple
graphs.

\begin{ex}  Consider the two  graphs $q$ and $G$ given in Figure
\ref{fig:graphSUBISO}. According to the label and neighborhood
inclusion condition, we can map vertex $u_1 \in V_q$ to $v_1 \in
V_{G}$ since (i) $l_{q}(u_1)$ = $l_{{G}}(v_1) = A$, and (ii)
$NL_q(u_1) = \{(B, Y), (B, Y)\} \subseteq  \{(A, X), (B, Y), (B,
Y)\} = NL_G(v_1)$.  $\blacksquare$
\end{ex}

Though the label and neighborhood inclusion condition is effective
in reducing  the search space, applying the inclusion test is
expensive especially for large size graphs with higher average
vertex degree. Below, we propose a new method to efficiently apply
the inclusion test. The method is based on the observation that many
vertices in the query or data graph share the same neighborhood. The
next example highlights this fact.

\begin{ex}  Consider the query graph $q$ and data graph $G$
given in Figure \ref{fig:graphSUBISO}. We have (1) In graph $G$:
$NL_G(v_1)$ = $NL_G(v_2)$ = $\{(A, X), (B, Y), (B, Y)\}$,
$NL_G(v_3)$ = $NL_G(v_5)$ = $\{(A, Y), (B, Z)\}$, and $NL_G(v_4)$ =
$\{(A, Y), (A, Y), (B, Z), (B, Z)\}$; (2) In query graph $q$:
$NL_q(u_1)$ = $\{(B, Y), (B, Y)\}$, and $NL_q(u_2)$ = $NL_q(u_3)$ =
$\{(A, Y), (B, Z)\}$.  $\blacksquare$
\end{ex}

Based on the above observation, we can reduce the cost of the
containment checks by caching most of the repeated computations, as
in the following steps:

\begin{enumerate}
\item Find the set of distinct labeled  neighborhoods for the
two graphs $q$  and $G$, denoted as $DLN_G$ and $DLN_q$, respectively.

\item Construct a bit matrix $M_{DLN} =(m_{ij})_{\alpha\beta}$
where $\alpha = |DLN_q|$ and $\beta=|DLN_G|$, to maintain the
inclusion relationship between distinct neighborhoods of $q$ and
$G$, that is, $m_{ij} = 1$ if $DLN_q[i] \subseteq DLN_G[j]$,
otherwise $m_{ij} = 0$.

\item For a graph $g$, where $g$ is $q$ or $G$, construct an array
of pointers  $P_g$ of size $|V_g|$, called position array, where
each slot $u$ holds the index of the vertex $u$ labeled neighborhood
at $DLN_g$.\\

\end{enumerate}

Now we can say that, for each $u \in V_q$ and $v \in V_G$, we have
$NL_q(u) \subseteq NL_G(v)$ iff $m_{{P_q(u)}{P_G(v)}} = 1$. Thus,
the test (ii) in label and neighborhood inclusion condition  can be
replaced by testing if $m_{{P_q(u)}{P_G(v)}} = 1$.\\

In subgraph search problem, for example, caching the repeated computations as
above is very useful since real graph data tend to share
commonality, that is, a vertex may appear in many data graphs. This
happens because the real data come from the same application domain.
Note that in the experiments,  subgraph search problem is used for testing {\tt Fast-ON} algorithm.

\begin{figure}[!htb]
\noindent \underline{\hspace*{10.5cm}}\\
\vspace{-0.10in}
\textbf{Algorithm:} $Order\_Vertices(V_q)$\\
\noindent \underline{\hspace*{10.5cm}}
\begin{tabbing}
\textbf{Input:} $V_q = \{u_1, u_2, \ldots, u_{|V_q|}\}$; \\
\textbf{Output:} An order of $V_q$, $V_q^\prime = \{u_1^\prime, u_2^\prime, \ldots, u_{|V_q|}^\prime\}$;\\\\
\bf{1:}\quad\=$V_q^\prime = \phi $;\\
\medskip
\bf{2:}\quad\={\bf for each} $u \in V_q$ {\bf do} calculate $deg(u)$;\\
\medskip
\bf{3:}\quad\=$u_1^\prime = u_k$, $k$ = argmax$_{u \in V_q} deg(u)$; \\
\medskip
\bf{4:}\quad\=Add $u_1^\prime$ to $V_q^\prime$ and remove $u_k$ from $V_q$;\\
\bf{5:}\quad\=\bf{for} $i = 2 \ldots |V_q|$\\
\bf{6:}\quad\quad\=$u_i^\prime = u_k$, $k$ = argmax$_{u \in V_q}$ $|\{(u, u^\prime) \in E_q: u^\prime \in V_q^\prime\}|$;\\
\bf{7:}\quad\quad\=Add $u_i^\prime$ to $V_q^\prime$ and remove $u_k$ from $V_q$;\\
\bf{8:}\quad\=\bf{return} $V_q^\prime$;\\
\noindent \underline{\hspace*{10.5cm}}
\end{tabbing}
 \caption{Ordering Query Vertices Algorithm}
\label{fig:OV}
\end{figure}

To speed up the search in {\tt Fast-ON}, we propose and ordering methodology of the query vertices as we show in the the following
optimization.
\subsubsection{Opt2: Ordering the query vertices}
\label{subsec:order}

This optimization is based on the observation  that the search order
in Ullman algorithm is random. It depends on the order of query
vertices imposed during input. This default ordering of $V_q$ can
possibly result in a search order that seriously slows down Ullman
Algorithm. Query vertices should be explored in the order that
facilitates  getting the utmost benefit of applying the third
condition. Unlike the QuickSI algorithm, our approach to order $V_q$ is to require the currently
processing query vertex to have high connectivity with the
previously explored ones, that is, suppose that $u_i\in V_q$ is the
currently processing vertex, then $u_i$ should have the higher
connectivity with $u_1, u_2, \ldots, u_{i-1}$ among the remaining
ones. Whereas, the first vertex to explore, i.e., $u_1$, is the one with maximum degree. This new
ordering forces false mapping to be discarded as early as possible
during the search, thus saving much of the time that Ullman
algorithm may take on false long partial mappings. Figure
\ref{fig:OV} outlines this idea.
\begin{figure}[!htb]
\noindent \underline{\hspace*{11cm}}\\
\vspace{-0.10in}
\textbf{Algorithm:} {\tt Fast-ON} $(q, G)$\\
\noindent \underline{\hspace*{11cm}}
\begin{tabbing}
\textbf{Input:} $q$: a query graph and $G$: a data graph. \\
\textbf{Output:} Boolean: $q$ is a subgraph of $G$.\\\\
{\bf Boolean} Test = {\tt FALSE}; \hspace{1.9cm} /* Global Variable */ \\\\
\bf{1:}\quad\=$V_q^\prime =  Order\_Vertices(V_q)$; \hspace{1.9cm} /* Opt2 */ \\
\bf{2:}\quad\=Construct $DLN_G, DLN_q$ and $M_{DLN}$; \\
\bf{3:}\quad\=Construct both $P_q$ and $P_G$; \\
\bf{4:}\quad\={\bf for} each $u \in V_q^\prime$ {\bf do}\\
\bf{5:}\quad\quad\=$Cand(u)$ = $\{v: v \in V_G, l_q(u) = l_G(v),$ and $m_{{P_q(u)}{P_G(v)}} = 1\}$; \hspace{0.2cm} /* Opt1 */\\
\bf{6:}\quad\=$Recursive\_Search(u_1)$;\\
\bf{7:}\quad\={\bf return} Test;\\\\
{\bf Procedure} $Recursive\_Search(u_i)$\\
\bf{1:}\quad\={\bf if} NOT Test {\bf then}\\
\bf{2:}\quad\quad\={\bf for} $v \in Cand(u_i)$ and $v$ is unmatched {\bf do} \hspace{0.3cm}/* Cond. 2 (Ullman)*/ \\
\bf{3:}\quad\quad\quad\={\bf if}  NOT  $Matchable(u_i, v)$ {\bf then} {\bf continue};\\
\bf{4:}\quad\quad\quad\=$f(u_i) = v$; $v$ = matched;\\
\bf{5:}\quad\quad\quad\={\bf if} $i < |V_q^\prime|$ {\bf then}\\
\bf{6:}\quad\quad\quad\quad\=$Recursive\_Search(u_{i+1})$;\\
\bf{7:}\quad\quad\quad\={\bf else} \\
\bf{8:}\quad\quad\quad\quad\=Test = {\tt TRUE}; \\
\bf{9:}\quad\quad\quad\quad\={\bf return}; \\
\bf{10:}\hspace{-.15cm}\quad\quad\quad\=$f(u_i) =$ {\tt NULL}; $v$ = unmatched; \hspace{0.4cm}/* Backtrack */ \\\\
{\bf Function} $Matchable(u_i, v)$ \hspace{2.1cm} /* Cond. 3 (Ullman)*/\\
\bf{1:}\quad\={\bf for} each $(u_i, u_j) \in E_q,$ $j < i$ {\bf do}\\
\bf{2:}\quad\quad\={\bf if} $(v, f(u_j)) \notin E_G$  {\bf then return} {\tt FALSE};\\
\bf{3:}\quad\={\bf return} {\tt TRUE};\\
\noindent \underline{\hspace*{11cm}}
\end{tabbing}
 \caption{{\tt Fast-ON} Algorithm}
\label{fig:ouralg1}
\end{figure}
\subsubsection{{\tt Fast-ON} Pseudocode}
Figure \ref{fig:ouralg1} outlines {\tt Fast-ON} algorithm. Line 1
applies the second optimization Opt2, whereas lines 2-5 outline the
first optimization Opt1.  In line 5, for each query vertex $u \in
V_q$, data graph vertices $v \in V_G$ that satisfy the modified
first condition are collected into a set called candidate set
$Cand(u)$. The procedure $Recursive\_Search$ matches $u_i$ over
$Cand(u_i)$ (line 5) and proceeds step-by-step by recursively
matching the subsequent vertex $u_{i+1}$ over $Cand(u_{i+1})$ (lines
6-7), or sets the Test variable to true value and returns if every vertex of $q$
has counterpart in $G$ (line 9). If $u_i$ exhausts all vertices in
$Cand(u_i)$  and still cannot find matching, Recursive\_Search
backtracks to the previous state for further exploration (line 11).
The procedure Matchable applies the third condition.\\

Note that according to Opt1, for each $u$, $Cand(u)$ is as  small as
possible. Consequently {\tt Fast-ON} explores much smaller space
than Ullman algorithm. Moreover, according to Opt2, false mappings
are discarded as early as possible, saving much of the computations
spent by Ullman algorithm.\\

\subsection{{\tt Fast-P} Algorithm } \label{sec:substructure}
The vertex-to-vertex matching used in Ullman and {\tt Fast-ON} is
time consuming specially when $N = |V_q|$ is large. Recall that $N$
represents the depth of the search tree. In this section, we
propose a new algorithm for subgraph isomorphism problem that uses
substructure correspondences instead of vertex correspondences to
reduce the depth of the search tree. Intuitively, if we index a set of
substructures of the data graph $G$, $S=\{s_1, s_2, \ldots\}$, such
that $s_i \subset G$, and answer subgraph isomorphism in a
structure-at-a-time manner by checking one-to-one correspondence on
query's substructures instead of query's vertices, we definitely
reduce the depth of the search space. In other words, we can
minimize the depth of the the search tree of Ullman algorithm by
matching a substructure per iteration. Applying this idea, two
challenges will arise which are as follows.
\begin{itemize}
\item {\emph{The First Challenge.}} Which kind of substructures will efficiently work?
\item {\emph{The Second Challenge.}} How  these substructures are extracted and used?\\
\end{itemize}

Regarding the first challenge, there are three kinds of substructures that can be indexed, that are
paths, trees, and graphs. We use paths for the following reasons:

\begin{enumerate}
\item Enumerating paths in a given graph $G$ is simple and easy
while enumerating general subgraphs or simply trees is quite
expensive.

\item Manipulating paths is much easier than that for general subgraphs.
For instance, the number of redundancies of every path's embedding
is at most two, while it could be much larger than two for general
subgraphs, which adds extra overhead for the case of general
subgraphs. The main cause of redundancy will be discussed in more
details below.
\end{enumerate}

The new algorithm, called {\tt Fast-P} ({\tt Fast} {\tt
P}ath-at-a-time manner algorithm), explores a tree-structured search
space considering all possible path-to-path mappings from $q$ to
$G$. Each path corresponding to a query path is, in fact, a local
match to its corresponding query path. If the query is subgraph
isomorphic to the data graph, then some of these local matches could
be combined together to produce a global match to the query. In what
follows, we show how paths are extracted and efficiently used in
{\tt Fast-P} (the second challenge).

\subsubsection{Path Enumeration and Encoding in {\tt Fast-P}}

Since the strategy of {\tt Fast-P} is based on path-to-path
matching, we first enumerate and index simple paths in the data
graph $G$. Usually, the number of paths in  $G$  is large. Thus, we
will use a path's size parameter, called $maxL$, to control the
number of indexed paths in $G$. We use ${\cal P}_G$ to denote the
set of simple paths of size up to  $maxL$ in a graph $G$.  To deal
with the issue of redundancy while path enumeration, we introduce
the following concepts.

\begin{definition} {\bf Reversed Path.}
Given a path $p = v_1 \rightsquigarrow  v_k$ in a graph $G$, its
reversed path is a path $v_k \rightsquigarrow v_1$ and denoted by
$p^{\tt r}$.  $\blacksquare$
\end{definition}

\begin{definition} {\bf (Non-)Iso Path.}
\label{def:noniso} A path $v_1 \rightsquigarrow  v_k$ in a graph $G$
is called an iso path if $l(v_i) = l(v_{k-i+1})$ and $l((v_i,
v_{i+1})) = l((v_{k-i}, v_{k-i+1}))$ $\forall$ $i = 1,2, \ldots,
k/2$, otherwise it is called a non-iso path.  $\blacksquare$
\end{definition}

\begin{figure}[h]
  \centerline{\psfig{figure=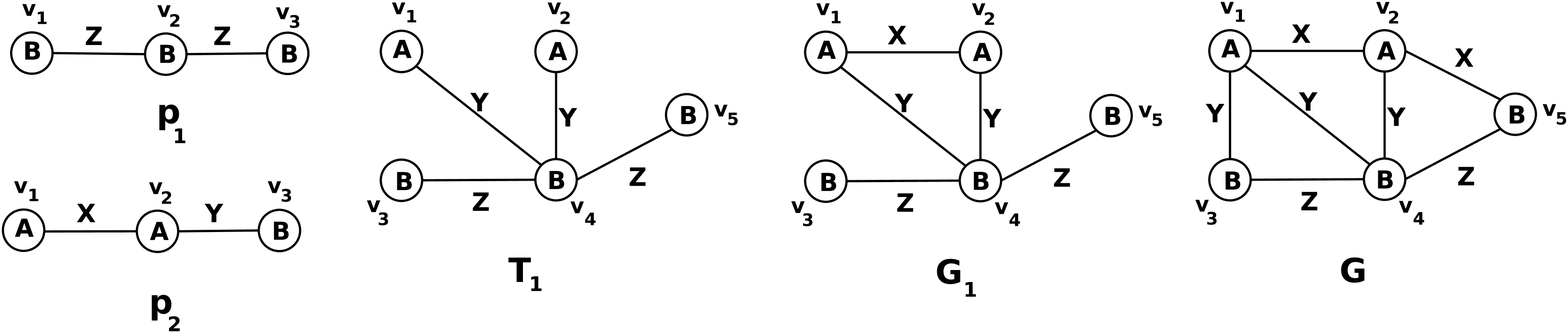,width=5.1in,height=2in} }
  \caption{Duplicated Embedding}
  \label{fig:duplicated-embedding}
\end{figure}

\begin{ex}
The path $p_1 = (v_1, v_2, v_3)$ in Figure
\ref{fig:duplicated-embedding} is called iso path since $l(v_1) =
l(v_3) = B$ and $l((v_1, v_2)) = l((v_2, v_3)) = Z$, while the path
$p_2$ is called non-iso path since $l(v_1) = A \neq B = l(v_3)$.
Finally, $p_1^r = (v_3, v_2, v_1)$.  $\blacksquare$
\end{ex}

\begin{lemma}
Every embedding of an iso path $p$ has two redundancies, $p$ and
$p^r$.
\end{lemma}


\begin{ex}
\label{ex:duplicate} Given the two paths $p_1$ and $p_2$, the
tree $T_1$, and the two graph $G_1$ and $G$  in Figure
\ref{fig:duplicated-embedding}. The iso path $p_1$ has two redundant
embeddings in $G$, that are, $\{v_3,v_4,v_5\}$ and $\{v_5,v_4,v_3\}$
while the non-iso path $p_2$ has only one redundant embedding, that
is, $\{v_2,v_1,v_4\}$. The tree $T_1$ and subgraph $G_1$ have four
redundant embeddings  in $G$, which are $\{v_1,v_2,v_3,v_4,v_5\}$,
$\{v_2,v_1,v_3,v_4,v_5\}$, $\{v_1,v_2,v_5,v_4,v_3\}$, and
$\{v_2,v_1,v_5,v_4,v_3\}$.   $\blacksquare$
\end{ex}

Storing and comparing paths would require a good representation of
path embeddings. To do so, consider the following concepts.

\begin{definition} {\bf Canonical Path.}
The code of a path $p = v_1 \rightsquigarrow  v_k$, denoted as
$code(p)$, is a sequence of vertex and edge labels in the following
order: $"l(v_1)l((v_1,v_2))l(v_2)\ldots l(v_{k-1})
l((v_{k-1},v_{k})) l(v_k)"$. The path $p$ is called canonical,
denoted $p^c$, if its code is the lexicographically minimum of
$code(p)$ and $code(p^r)$.  $\blacksquare$
\end{definition}

\begin{corollary}
Every iso path $p$ is canonical.
\end{corollary}

{\sc proof:} This is because $code(p) = code(p^r)$.   $\blacksquare$

\begin{ex}
Consider the two paths $p_1$ and $p_2$ in Figure
\ref{fig:duplicated-embedding},  we have $p_1$ is canonical since it
is iso path, and the path $p_2$ is canonical since $code(p_2)$ =
$"AXAYB" \leq "BYAXA" = code(p_2^c)$.  $\blacksquare$
\end{ex}

\subsubsection{Path Matching in {\tt Fast-P}}

Usually, the number of paths in a query that are candidates  for
matching is much larger than the number of vertices, i.e., $|{\cal
P}_q| \gg |V_q|$. Thus, for {\tt Fast-P} to be effective, the number
of query's paths used for matching should be less than the number of
query's vertices. Considering disjoint paths of size up to $maxL$,
denoted as $DP_{maxL}(q)$, which cover the query, is a key step
toward reaching this objective. Disjoint paths are defined as
follows.

\begin{definition} {\bf Disjoint Paths.}
Distinct paths in a graph $q$ are called {\em disjoint} if they are
edge disjoint,  but not necessarily node disjoint.  $\blacksquare$
\end{definition}

\begin{ex}
Suppose that the graph $G$ in Figure \ref{fig:duplicated-embedding} is our query $q$, and set $maxL =
2$. There are 21 paths in ${\cal P}_q$ given as: ${\cal P}_q =$
{\large\{}
$\{1,2,4\},\{1,2,5\},\{1,3,4\},\{1,4,2\},\{1,4,3\},\{1,4,5\},\{2,1,3\},\{2,1,4\},\{2,4,3\},\{2,4,5\},\\\{2,5,4\},\{3,1,4\},\{3,4,5\},\{4,2,5\},\{1,2\},\{1,3\},\{1,4\},\{2,4\},\{2,5\},\{3,4\},\{4,5\}
${\large \}}. The following paths are disjoint paths covering $q$, $DP_{2}(q)$ = {\large
\{}$\{5,2,4\},\{2,1,4\},\{3,4,5\},\{1,3\}${\large \}}. Compare
$|DP_{2}(q)| = 4$ with $|V_q| = 7$, we can save three call with {\tt
Fast-P}.  $\blacksquare$

\end{ex}

Thus, the total search space of {\tt Fast-P} is given by the product
$\prod^{|DP_{maxL}(q)|}_{i=1} Cand(p_i)$, where $Cand(p_i)$ =
$\{p^\prime \in {\cal P}_G: code(p^{\prime^c}) = code(p_i^c)\}$ is
the set of graph paths that match a query path $p_i$ . To optimize
{\tt Fast-P}, query paths should be chosen such that
$|DP_{maxL}(q)|$ and $|Cand(p_i)|$ are minimized. The first
optimization we introduce, called Opt1, minimizes $|DP_{maxL}(q)|$.
Another optimization called Opt2 is used to minimize the set of
matching candidates $|Cand(p_i)|$ for each query path $p_i$.
Finally, to speed up the search in {\tt Fast-P}, we propose an
ordering of the query paths to force false mappings to be discarded
as early as possible during the search. This ordering is presented in the third optimization, called  Opt3.\\

\begin{figure}[!htb]
\noindent \underline{\hspace*{11.5cm}}\\
\vspace{-0.10in}
\noindent\textbf{Algorithm:} {\tt Cover} $(q, P_q)$\\
\noindent \underline{\hspace*{11.5cm}}\\
Input: ${\cal P}_q$: $q$'s simple paths of size up to $maxL$-edges; \\
Output: $DP_{maxL}(q)$: disjoint paths covering $q$, initialized empty;\\
\begin{tabbing}
\bf{1:}\quad\=Sort ${\cal P}_q$ in decreasing order based on path size;\\
\bf{2:}\>$DP_{maxL}(q) = \{\}$;\\
\bf{3:}\> $q^\prime = q$;\\
\bf{4:}\>\textbf{for} each $p \in {\cal P}_q$ \textbf{do}\\
\bf{5:}\>\quad\=\textbf{if} $p \subseteq q^\prime$ and  $q^\prime \setminus p$ is connected \textbf{then}\\
\bf{6:}\>\>\quad\= Remove $p$ from $q^\prime$; \\
\bf{7:}\>\>\>$DP_{maxL}(q) = DP_{maxL}(q) \cup \{p\}$;\\
\bf{8:}\>\>\textbf{if} $q^\prime$ is empty graph \textbf{then} \\
\bf{9:}\>\>\>\bf{break};\\
\bf{10:}\>\textbf{return} $DP_{maxL}(q)$;\\
\noindent \underline{\hspace*{11.5cm}}
\end{tabbing}
\caption{An Algorithm to Find a  Cover of $q$} \label{fig:cover}
\end{figure}

\subsubsection{Opt1: Minimizing $|DP_{maxL}(q)|$}
\label{subsec:minimizeDP}

For a given query graph, there are multiple disjoint path
decompositions. Some are compact and the others are not. The
algorithm in Figure \ref{fig:cover}  finds a compact set of disjoint
paths that cover $q$. The algorithm works as follows. Given the set
of all limited-size, simple paths ${\cal P}_q$ generated from the query
$q$. ${\cal P}_q$ is processed in descending order of path size. For each
encountered path $p \in {\cal P}_q$, we check if removing $p$ from the
query disconnects it or not. If so, i.e., the resulting graph is
disconnected, $p$ is not considered and the search continue for
another one. If, on the other hand, the resulting graph still
connected, $p$ is selected to be in the cover and removed from the
query. Theorem ~\ref{thm:compact} shows that the selected paths
$DP(q)$ are disjoint, and if $maxL = 2$, then $DP(q)$ is compact.

\begin{theorem}\label{thm:compact}
Given ${\cal P}_q$, the set of $q$ simple paths of size up to
$maxL$-edges. The set $DP_{maxL}(q)$  returned by the algorithm in
Figure \ref{fig:cover} is the set of disjoint paths covering $q$. If
$maxL = 2$, then $DP_{maxL}(q)$ is compact.
\end{theorem}

{\sc proof:} A path of the largest length $p \in {\cal P}_q$ is
inserted into $DP_{maxL}(q)$ (line 7)  and removed from $q^\prime$
(line 6) if  it fully exists in $q^\prime$, i.e., if $p \subseteq
q^\prime$ (line 5). This guarantees that all chosen paths do not
share any edge, i.e., they are disjoint.

Suppose that $DP_{maxL}(q)$ is not compact and $maxL = 2$. Then,
there exist at least  two 1-edge paths $p_1$ and $p_2$ in
$DP_{maxL}(q)$ such that the path $p = p_1 \cup p_2$ is not chosen
by the algorithm. Since $p_1 \subseteq q^\prime$ and $p_2\subseteq
q^\prime$, then the only reason to not choose $p$ is that $p$
disconnects $q^\prime$. On the other hand, since removing $p_1$ or
$p_2$ leaves $q^\prime$  connected, then removing $p$ also leaves
$q^\prime$ connected, i.e., $p$ should have been chosen, a
contradiction.  $\blacksquare$\\

According to Theorem \ref{thm:compact}, if we set $maxL = 2$, then
$DP_{maxL}(q)$ is  compact and we have two cases with respect to the
number of edges in $q$ as follows.

\begin{itemize}
\item If $|E_q|$ is even then $DP_{maxL}(q)$
contains $\lfloor|E_q|/2\rfloor$  paths of size 2 (i.e.
$|DP_{maxL}(q)| = \lfloor|E_q|/2\rfloor$).

\item If $|E_q|$ is odd then $DP_{maxL}(q)$
contains $\lfloor|E_q|/2\rfloor$ paths of size 2  and one path of
size 1 (i.e. $|DP_{maxL}(q)| = \lfloor|E_q|/2\rfloor + 1$).
\end{itemize}

\begin{ex}
Consider the  query $q$ in Figure \ref{fig:database} and set $maxL =
2$. Since $|E_q| = 7$ is odd then $|DP_{2}(q)| = \lfloor7/2\rfloor +
1 = 4$. The following disjoint paths are generated using the
algorithm in Figure \ref{fig:cover}:  $DP_{2}(q)$ = ${\large
\{}\{3,1,2\},\{1,4,2\},\\\{5,2,3\},\{4,5\}{\large \}}$. The size of
$DP_{2}(q)$ is optimal.  $\blacksquare$
\end{ex}

Unfortunately, there is a tradeoff between the number of calls
(depth of the search space) in {\tt Fast-P} and the $maxL$ used. For
instance, suppose the query $q$ is a complete graph such that $|V_q|
= 7$ then $q$ has $|E_q| = (|V_q|.(|V_q|-1))/2 = 21$ edges. Choosing
$maxL = 1$, Algorithm {\tt Cover} will produce $|DP_{1}(q)| = 21$
disjoint paths, i.e., the number of edges in $q$.  Setting $maxL =
2$, we still have 11 disjoint paths that cover $q$. Comparing with
$|V_q| = 7$, substructure matching of paths of size 2 is not
effective in this case.\\

To guarantees a higher efficiency than that of vertex-at-a-time
approaches, $maxL$ must be chosen according to the following
equation.

\begin{equation}
\label{equ:maxLL} |E_q|/|V_q| < maxL
\end{equation}

To set equation \ref{equ:maxLL} in terms of graph density, where the
density of query $q$ is defined as $d_q =
2.|E_q|/(|V_q|.(|V_q|-1))$. Then equation \ref{equ:maxLL} will be
given as:
\begin{equation}
d_q < 2.maxL/(|V_q|-1)
\end{equation}

This equation shows the role that query density plays in the
performance of {\tt Fast-P}. Dense queries require higher $maxL$.
Fortunately, the real data and the queries are always sparse graphs.

\begin{ex}
Consider the query $q$ in Figure \ref{fig:graphSUBISO}. Since $|V_q|
= 3$ and $|E_q| = 3$, setting $maxL = 2$ will make {\tt Fast-P}
faster than Fast-{\tt ON}. $\blacksquare$
\end{ex}



\subsubsection{Opt2: Minimizing $|Cand(p_i)|$}
\label{subsec:minimizeCP} For each query path $p$, $Cand(p)$ is
guaranteed to be smaller than $\prod_{v_i \in p} Cand(v_i)$. This is
because vertex connections are already considered in the paths. For
instance, consider a query path $p = (v_1, v_2, v_3)$, and given
$Cand(v_1)$, $Cand(v_2)$, and $Cand(v_3)$. There are $Cand(v_1)
\times Cand(v_2) \times Cand(v_3)$ combinations to be considered in
any vertex-to-vertex manner algorithm. On the other hand, the number
$|Cand(p)|$ is much smaller than the previous product, since all
paths connecting the vertices in $Cand(v_1)$, $Cand(v_2)$, and
$Cand(v_2)$ are the only considered ones. Hereafter, we optimize
$Cand(p)$, i.e.,  reduce the candidate set of each  path $p \in
DP_{maxL}(q)$ more than ever, by utilizing the neighborhood
labels of all vertices in $p$.\\

The next theorem presents the necessary condition required by any
data graph path $p^\prime \in Cand(p)$ to share in any subgraph
isomorphism between $q$ and the data graph $G$.

\begin{theorem}
\label{thm:containmentneighborpath} If the query graph $q$ is
subgraph isomorphic to the data graph $G$, then  for any  $p^\prime
\in Cand(p)$ sharing in the isomorphism, $p^c = (u_1, \ldots, u_k)$
and $p^{\prime^c} = (v_1, \ldots, v_k)$ must satisfy

\begin{enumerate}
\item $NL_q(u_i) \subseteq NL_G(v_i)$ \hspace{3.1cm}$\forall$ $i =1,\ldots ,k$, or
\item $NL_q(u_i) \subseteq NL_G(v_{k - i + 1})$ \hspace{2.3cm}$\forall$ $i =1,\ldots
,k$.
$\blacksquare$
\end{enumerate}
\end{theorem}

The previous theorem presents the necessary condition required for a
data graph path $p^\prime$ to be included in $Cand(p)$, $p\in
DP_{maxL}(q)$. Applying this condition while constructing $Cand(p)$
would minimize $Cand(p)$ and cut down the search space of {\tt
Fast-P}.

\begin{corollary}
In the case of non-iso path $p$, the first test is sufficient. $\blacksquare$
\end{corollary}

To efficiently apply the inclusion tests  in {\tt Fast-P} algorithm,
we construct a  bit matrix similar to that is used with {\tt
Fast-ON}, $M_{DLN} =(m_{ij})_{\alpha\beta}$ (where $\alpha =
|DLN_q|$ and $\beta=|DLN_G|$) and the same two pointers $P_q$ and
$P_G$  as in the {\tt Fast-ON} algorithm.  The two tests in Theorem
\ref{thm:containmentneighborpath} are replaced by the following two
tests:
\begin{enumerate}
\item $m_{{P_q(u_i)}{P_G(v_i)}} = 1$ \hspace{2.65cm}$\forall$ $i =1,\ldots
,k$, or
\item $m_{{P_q(u_i)}{P_G(v_{k - i + 1})}} = 1$ \hspace{2cm}$\forall$ $i =1,\ldots
,k$.
\end{enumerate}

\subsubsection{Opt3: Ordering $DP_{maxL}(q)$}
\label{subsec:orderingDP} Although $Cand(p_i)$ is minimized for each
$p_i \in DP_{maxL}(q)$ based on Opt2, the  search order of the paths
in  $DP_{maxL}(q)$  is random, and can seriously slow down the
algorithm. Query disjoint paths $DP_{maxL}(q)$ should be explored in
the order that excludes false local matches  of each path $p_i \in
DP_{maxL}(q)$ as early as possible, saving much of the time that may
be taken on false long partial mappings.  A local match of path
$p_i$  is false if it does not satisfy the preservation of
structural connectivity. When we maximize the node overlapping of a
currently processing query disjoint path $p_i \in DP_{maxL}(q)$ with
the previously explored ones ($p_1, ..., p_{i-1}$), we, in fact,
maximize the connectivity among  $p_i$ and the previously explored
ones ($p_1, ..., p_{i-1}$), and thus increase the likelihood that
false local matches are detected early. Thus, we adopt an ordering
of $DP_{maxL}(q) = \{p_1, p_2, \ldots, p_{|DP_{maxL}(q)|}\}$, such
that the node overlapping of $V_{p_i}$ is maximized with $\cup_{j <
i} V_{p_j}$. And, the first path $p_1$ is chosen such that
$\sum_{u\in V_{p_1}} freq(u)$ is maximum, where $freq(u)$ is the
frequency of the node $u$ with respect to $DP_{maxL}(q)$. Figure
\ref{fig:order-path} outlines the idea.\\

\begin{figure}[h]
\noindent \underline{\hspace*{11cm}}\\
\vspace{-0.10in}
\noindent\textbf{Algorithm:} {\tt Order}$(V_q, DP_{maxL}(q))$\\
\noindent \underline{\hspace*{11cm}}\\
{\bf Input:} $DP_{maxL}(q) = \{p_1, p_2,\ldots, p_{|DP_{maxL}(q)|}\}$; \\
{\bf Output:} An order of $DP_{maxL}(q) = \{p^{\prime}_1,
p^{\prime}_2,\ldots, p^{\prime}_{|DP_{maxL}(q)|}\}$;
\begin{tabbing}
{\bf 1:}\quad\=\textsf{for} each $u \in V_q$ \textsf{do} calculate $freq(u)$;\\
{\bf 2:}\>$p^{\prime}_1 = p_k$, $k =$ argmax$_{p \in DP_{maxL}(q)}  \sum_{u\in V_{p}} freq(u)$;\\
{\bf 3:}\>$DP_{maxL}(q) = DP_{maxL}(q) \setminus \{p^{\prime}_1\}$; \\
{\bf 4:}\>$newDP_{maxL}(q) = \{p^{\prime}_1\}$; \\
{\bf 5:}\>$V = V_{p^{\prime}_1}$;\\
{\bf 6:}\>\textsf{for}  $i = 2\ldots (|DP_{maxL}(q)|-1)$ \textsf{do}\\
{\bf 7:}\>\quad\= $p^{\prime}_i = p_k$,  $k =$ argmax$_{p \in DP_{maxL}(q)} |V_p \cap V|$;\\
{\bf 8:}\>\>$DP_{maxL}(q) = DP_{maxL}(q) \setminus \{p^{\prime}_i\}$;\\
{\bf 9:}\>\>$newDP_{maxL}(q) = newDP_{maxL}(q) \cup \{p^{\prime}_i\}$; \\
{\bf 10:}\>\>$V = V \cup V_{p^{\prime}_i}$; \\
{\bf 11:}\>\textsf{return} $newDP_{maxL}(q)$;\\
\noindent \underline{\hspace*{11cm}}
\end{tabbing}
\caption{Algorithm for Ordering $DP_{maxL}(q)$}
\label{fig:order-path}
\end{figure}
\begin{figure}[h]
\noindent \underline{\hspace*{11cm}}\\
\vspace{-0.10in}
\textbf{Algorithm:} {\tt Fast-P(q, G)}\\
\noindent \underline{\hspace*{11cm}}
\begin{tabbing}
\textbf{Input:} $q$: a query graph and $G$: a data graph. \\
\textbf{Output:} Boolean: $q$ is a subgraph of $G$.\\\\
{\bf Boolean} Test = {\tt FALSE}; \hspace{1.9cm} /* Global Variable */ \\\\
\bf{1:}\quad\={\bf for each} $u \in V_q$ {\bf do}\\
\bf{2:}\quad\quad\=$u.Count = 0$ \\
\bf{3:}\quad\quad\=$h[u] =$ {\tt NULL} \\
\bf{4:}\quad\={\bf for each} $v \in V_G$ {\bf do}\\
\bf{5:}\quad\quad\=$v.Count^{\prime} = 0$ \\
\bf{6:}\quad\=${\cal P}_q = \{p \subseteq q: p$ is a simple path $\wedge$ $|p| \leq maxL\}$;\\
\bf{7:}\quad\=${\cal P}_G = \{p^{\prime} \subseteq G: p^{\prime}$ is a simple path $\wedge$ $|p^{\prime}| \leq maxL\}$;\\
\bf{8:}\quad\=$DP_{maxL}(q) =  Cover(q, {\cal P}_{q})$; \hspace{1.5cm} /* Opt1 */ \\
\bf{9:}\quad\=$DP^{*}_{maxL}(q)=Order(V_q, DP_{maxL}(q))$;  \hspace{1.5cm}  /* Opt3 */\\
\bf{10:}\hspace{-.15cm}\quad\=Construct $DLN_G, DLN_q$ and $M_{DLN}$; \\
\bf{11:}\hspace{-.15cm}\quad\=Construct both $P_q$ and $P_G$; \\
\bf{12:}\hspace{-.15cm}\quad\={\bf for} each $p \in DP^{*}_{maxL}(q)$ {\bf do}\\
\bf{13:}\hspace{-.15cm}\quad\quad\={\bf if} $p$ is iso labeled path\\
\bf{14:}\hspace{-.15cm}\quad\quad\quad\=$Cand(p)$ = $\{p^{\prime} \cup {p^{\prime}}^r: p^{\prime} \in {\cal P}_G,$  $p^{\prime}$ satisfies Theorem \ref{thm:containmentneighborpath} \} /* Opt2 */\\
\bf{15:}\hspace{-.15cm}\quad\quad\={\bf else}\\
\bf{16:}\hspace{-.15cm}\quad\quad\quad\=$Cand(p)$ = $\{p^{\prime}: p^{\prime} \in {\cal P}_G,$  $p^{\prime}$ satisfies Theorem \ref{thm:containmentneighborpath} \} /* Opt2 */\\
\bf{17:}\hspace{-.15cm}\quad\quad\=$f(p) =$ {\tt NULL}\\
\bf{18:}\hspace{-.15cm}\quad\=$Recursive\_Search(p_1)$; \hspace{1.5cm} /* $p_1$ is the first path in $DP^{*}_{maxL}(q)$ */\\
\bf{19:}\hspace{-.15cm}\quad\={\bf return} Test;\\
\noindent \underline{\hspace*{11cm}}
\end{tabbing}
 \caption{Fast-P Algorithm}
\label{fig:ouralg23}
\end{figure}
\begin{figure}[h]
\noindent \underline{\hspace*{11cm}}\\
\vspace{-0.10in}
\textbf{Algorithm:} {\tt Recursive\_Search($p_i$)}\\
\noindent \underline{\hspace*{11cm}}
\begin{tabbing}
\bf{1:}\quad\={\bf if} NOT Test {\bf then}\\
\bf{2:}\quad\quad\={\bf for} $p^{\prime} \in Cand(p_i)$ and $p^{\prime}$ is unmatched {\bf do} \hspace{0.3cm}\\
\bf{3:}\quad\quad\quad\={\bf if}  NOT  $Matchable(p_i, p^{\prime})$ {\bf then} {\bf continue};\\
\bf{4:}\quad\quad\quad\=$f(p_i) = p^{\prime}$; $p^{\prime}$ = matched;\\
\bf{5:}\quad\quad\quad\={\bf if} $i < |DP^{*}_{maxL}(q)|$ {\bf then}\\
\bf{6:}\quad\quad\quad\quad\=$Recursive\_Search(p_{i+1})$;\\
\bf{7:}\quad\quad\quad\={\bf else} \\
\bf{8:}\quad\quad\quad\quad\=Test = {\tt TRUE}; \\
\bf{9:}\quad\quad\quad\quad\={\bf return}; \\
\bf{10:}\hspace{-.15cm}\quad\quad\quad\=$f(p_i) =$ {\tt NULL}; $p^{\prime}$ = unmatched; \hspace{0.4cm}/* Backtrack */ \\
\bf{11:}\hspace{-.15cm}\quad\quad\quad\={\bf for} $j = 1$ to $|V_{p_i}|$ {\bf do} \\
\bf{12:}\hspace{-.15cm}\quad\quad\quad\quad\=Set $u$ as $j$-th vertex in ${p_i}^c$ and  $v$ as $j$-th vertex in ${p^\prime}^c$\\
\bf{13:}\hspace{-.15cm}\quad\quad\quad\quad\= $u.Count = u.Count - 1$, $v.Count^\prime = v.Count^\prime - 1$, and $h[u] =$ {\tt NULL};\\\\
{\bf Function} $Matchable(p_i, p^\prime)$ \hspace{2.1cm} \\
\bf{1:}\quad\={\bf for} $j = 1$ to $|V_{p_i}|$ {\bf do}\\
\bf{2:}\quad\quad\=Set $u$ as $j$-th vertex in ${p_i}^c$ and  $v$ as $j$-th vertex in ${p^\prime}^c$\\
\bf{3:}\quad\quad\={\bf if} $((u.Count$ $\neq$ $0$ $||$ $v.Count^\prime$ $\neq$ $0)$ $\wedge$ $h[u] \neq v$) \\
\bf{4:}\quad\quad\quad\={\bf return} {\tt FALSE};\\
\bf{5:}\quad\={\bf for} $j = 1$ to $|V_{p_i}|$ {\bf do}\\
\bf{6:}\quad\quad\=Set $u$ as $j$-th vertex in ${p_i}^c$ and  $v$ as $j$-th vertex in ${p^\prime}^c$\\
\bf{7:}\quad\quad\= $u.Count = u.Count + 1$, $v.Count^\prime = v.Count^\prime + 1$, and $h[u] = v$\\
\bf{8:}\quad\={\bf return} {\tt TRUE};\\
\noindent \underline{\hspace*{11cm}}
\end{tabbing}
 \caption{Fast-P Algorithm (Continued)}
\label{fig:ouralg24}
\end{figure}

\subsubsection{{\tt Fast-P} Pseudocode}
The pseudocode of  {\tt Fast-P} is similar  to that of {\tt Fast-ON}
algorithm, except that paths are used instead of vertices. Figures
\ref{fig:ouralg23} and \ref{fig:ouralg24} outline the pseudocode of {\tt
Fast-P} algorithm. The main difference between {\tt Fast-P} and {\tt
Fast-P} codes  is that a query vertex has only one image at a time
in {\tt Fast-ON}. But it could have more than one image in {\tt
Fast-P}. This is because the query vertex could appear in many query
disjoint paths, and thus it has different images in the different
candidate paths of the data graph. To overcome this in {\tt Fast-P},
we combine candidate paths only if these  paths have  the same
images of a given vertex. To implement this, two counters are used
in {\tt Fast-P}, one for each vertex $u \in V_q$ denoted by
$u.Count$ and the other for each vertex $v \in V_G$, denoted by
$v.Count^\prime$. If $u \in V_q$ is mapped to vertex $v \in V_G$,
denoted as $h[u] = v$, then we increment $u.Count$ and
$v.Count^\prime$ (Lines 5-7 in function Matchable [Figure \ref{fig:ouralg24}]) by one and in the
backtracking step, we decrement one from $u.Count$ and
$v.Count^\prime$ (Lines 11-13 in
Recursive\_Search($p_i$) algorithm [Figure \ref{fig:ouralg24}]).\\

Regarding Figure \ref{fig:ouralg23}, Lines 1-5 initialize for each vertex query and
graph vertex its counter, and initialize for each vertex $u \in V_q$
its mapping by $0$ ($h[u]=0$). Lines 6-7 enumerate all simple paths
of size up to $maxL$ in $q$ and $G$ respectively. Line 8 applies the
first optimization (Opt1), whereas line 9  outlines the second
optimization (Opt2). Lines 10-16 apply the third optimization
(Opt3). Line 17 initializes the mapping ($f$)
that maps each path in $DP^{*}_{maxL}(q)$ to  {\tt NULL}.\\

The procedure $Recursive\_Search$ (Figure \ref{fig:ouralg24}) matches
a previously unmatched $p_i \in DP^{*}_{maxL}(q)$ over $Cand(p_i)$,
and proceeds step-by-step by recursively matching the subsequent
path $p_{i+1}$ over $Cand(p_{i+1})$ (lines 6-7), or sets Test to true
value (line 8) and returns if every path $p_i \in DP^{*}_{maxL}(q)$
has counterpart in ${\cal P}_G$ (line 9). If $p_i$ exhausts all
paths in $Cand(p_i)$ and still cannot find matching,
$Recursive\_Search$ backtracks to the previous state for further
exploration (lines 10-13). In function $Matchable$ (Figure\ref{fig:ouralg24}),
$p_i \in DP^{*}_{maxL}(q)$ is not mapped to
$p^\prime$ in $Cand(p_i)$, if for each $j$ such that $1 \leq j \leq
|V_{p_i}|$, the mapping $h[u] = v$ (where
$u$ is $j$-th vertex in ${p_i}^c$ and  $v$ is $j$-th vertex in ${p^\prime}^c$)
is not satisfied. In this case the function
Matchable return {\tt FALSE},
otherwise the function Matchable return {\tt TRUE}.

\section{Experimental Evaluation}
\label{sec:experiments} The experimental evaluation of
the two algorithms,  {\tt Fast-ON} and {\tt Fast-P}, are made
using PC with Intel 3GHz dual  Core CPU and 4G main memory and
running  Linux. The algorithms were implemented in standard C++ with
STL library support and compiled with GNU GCC. To make the time
measurements more reliable, no other applications were running on
the machine while doing the experiments. In experiments, we consider
vertex/edge labeled graphs  and  vertex
labeled graphs.\\

The rest of this chapter is organized as follows.  In Section
\ref{sec:datasetRS}, we present the datasets that are used in our
evaluation. Effects of optimization methods are presented in Section
\ref{sec:opt-ON-P}. Finally, in the reminding sections, we present
experimental results of the two algorithm ({\tt Fast-ON} and {\tt
Fast-P}).

\subsection{Datasets}
\label{sec:datasetRS}
\subsubsection{Real Dataset}
{\bf AIDS\_10K.} The first real dataset, referred to as AIDS\_10k,
consists of 10,000 graphs that are randomly drawn from the AIDS
Antiviral screen database \footnote{http://dtp.nci.gov/.}. These graphs have
25 vertices  and 27 edges on average. There are totally 62 distinct
vertex labels in the dataset but the majority of these labels are C,
O and N. The total number of distinct edge labels is 3.\\\\
{\bf Chem\_1M.} In order to study the scalability of {\tt Fast-ON} and {\tt Fast-P}
against different dataset size, we  use a large real chemical
compound dataset, referred to as Chem\_1M. Chem\_1M is a subset of
the PubChem database (ftp://ftp.ncbi.nlm.nih.gov/pubchem/), and
consists of one million graphs. Chem\_1M has 23.98 vertices and
25.76 edges on average. The number of distinct vertex and distinct
edge labels are 81 and 3, respectively. For this study, we derive
subsets from Chem\_1M, each one consists of N graphs and called
Chem\_N dataset. Note that the  Chem\_1M is the same as that used in
~\cite{iGraph}.

\subsubsection{Synthetic Datasets}
The synthetic datasets are generated  using the synthetic graph data
generator GraphGen~\cite{FG-Index}. The generator allows us to
specify various parameters such as the average graph density D,
graph size E and the number of distinct vertex/edge labels L. For
example, Syn10K.E30.D5.L50 means that it contains 10,000 graph; the
average size of each graph is 30;  the density of each graph is 0.5;
and the number of distinct vertex/edge labels is 50. Five synthetic
datasets with varying parameter values are used in experiments in
order to see performance changes with varying parameter values
(Syn10K.E30.D3.L50, Syn10K.E30.D5.L50, Syn10K.E30.D7.L50,
Syn10K.E30.D5.L80 and Syn10K.E30.D5.L20). Note that all the previous
five synthetic datasets are dense dataset and are the same as in
\cite{iGraph}.  Also, we get another synthetic dataset from CT-index
\cite{CT-Index}. This dataset is sparse dataset and we denote it by
SynCT\_10K.

\subsubsection{Query Sets} For each dataset (real or  synthetic),
there are six query sets Q4, Q8, Q12, Q16, Q20 and Q24. Each Qi
consists of 1000 queries, each of which of size $i$. For AIDS\_10K,
Chem\_1M, and the previous five synthetic datasets, we adopt the
query set from \cite{iGraph}. For SynCT\_10K, we adopt
the query set from \cite{CT-Index}.

\subsection{Performance of Subgraph Checking Algorithms}
\subsubsection{Effects of Optimizations}
\label{sec:opt-ON-P}
In this section, we show the effect of each
optimization on the performance of {\tt Fast-ON} and {\tt Fast-P}
algorithms.
\begin{itemize}
\item {\emph{Effects of Optimizations in {\tt Fast-ON} Algorithm}}\\
There are two optimizations, called Opt1 and Opt2, introduced in
{\tt Fast-ON}.  In this experiment, we show the effect of each
optimization independently, and the effect of them combined, on the
performance of {\tt Fast-ON}. For this purpose, we implemented three
versions of {\tt Fast-ON}, namely, {\tt Fast-O} that uses only the
first optimization Opt1, {\tt Fast-N} that uses only the second
optimization Opt2, and {\tt Fast-ON} that uses both of the two
optimizations.
\begin{figure}[h]
\centering
\epsfig{file=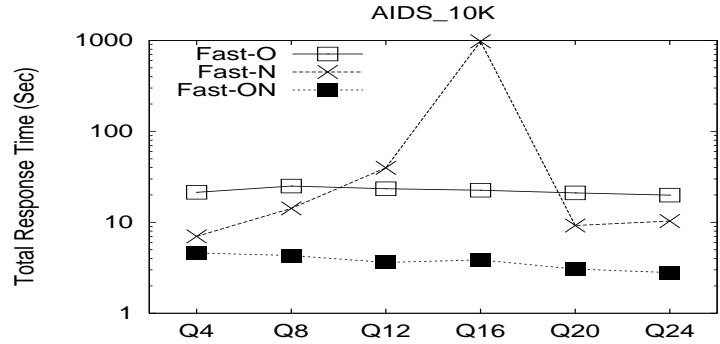,width=1.9in,height=4in,angle=-90}
\caption{Effects of Optimizations in {\tt Fast-ON} Algorithm}
\label{fig:effectoptsfast-on}
\end{figure}
Figure \ref{fig:effectoptsfast-on} plots the results obtained by
running the three versions on AIDS\_10K for the different query
sets. The figure shows that {\tt Fast-N} is faster than {\tt Fast-O}
except for Q12 and Q16, where {\tt Fast-O} shows the best
performance. In addition to its influence on speed, the first
optimization makes the algorithm less sensitive to query size.  {\tt
Fast-ON} shows the best performance, it outperforms both {\tt
Fast-O} and {\tt Fast-N}. This result confirms the fact that the two
optimizations are neither independent nor conflicting, but they are
complementary to each other.

\item {\emph{Effects of Optimizations in {\tt Fast-P} Algorithm}}\\ In
{\tt Fast-P} Algorithm, there are three optimizations, called Opt1,
Opt2, and Opt3. In this experiment, we show the effect of each
optimization independently, and the effect of them combined, on the
performance of {\tt Fast-P}.
\begin{figure}[h]
\centering
\subfigure[]{\epsfig{file=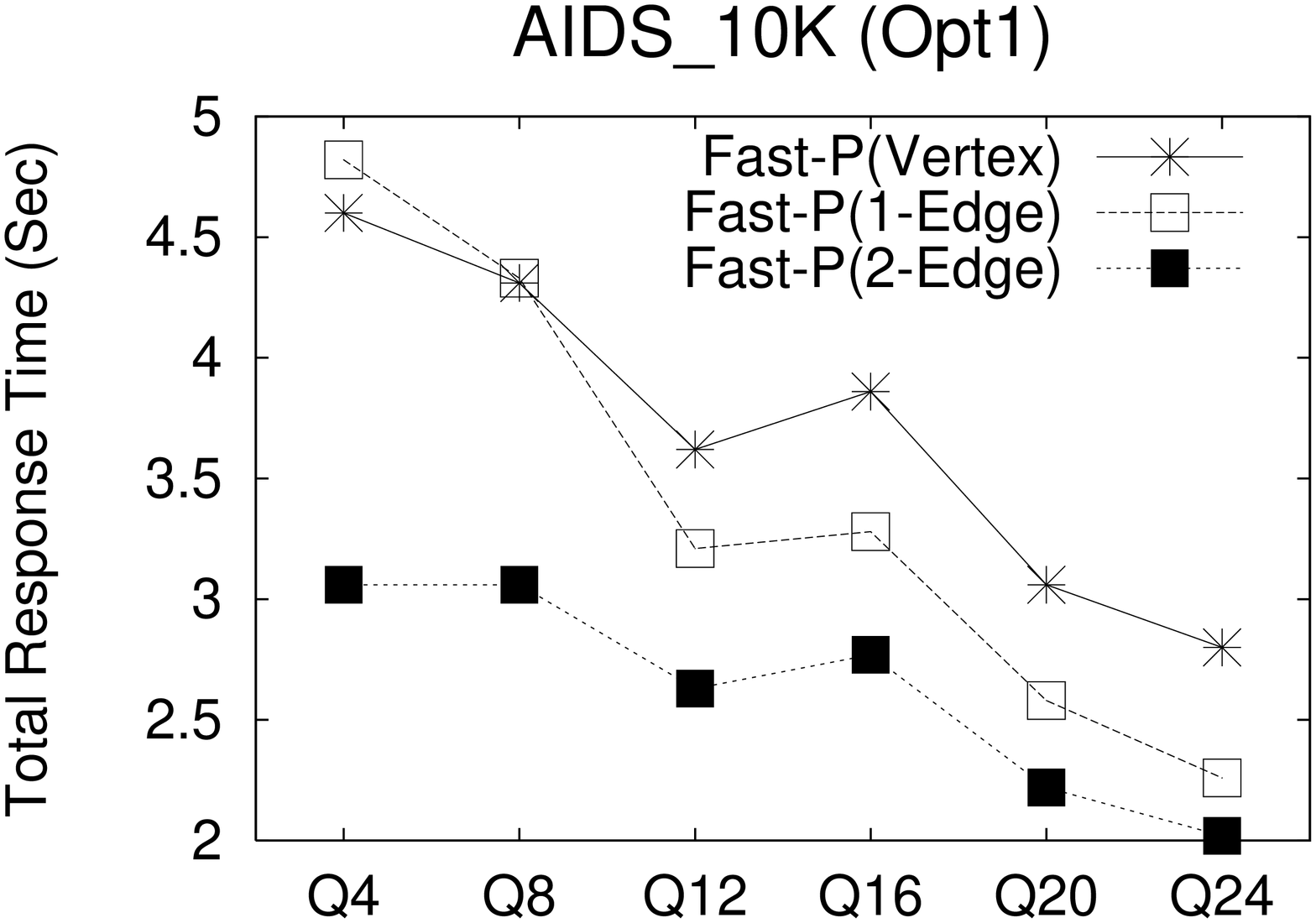,width=1.9in,height=4in,angle=-90}}
\subfigure[]{\epsfig{file=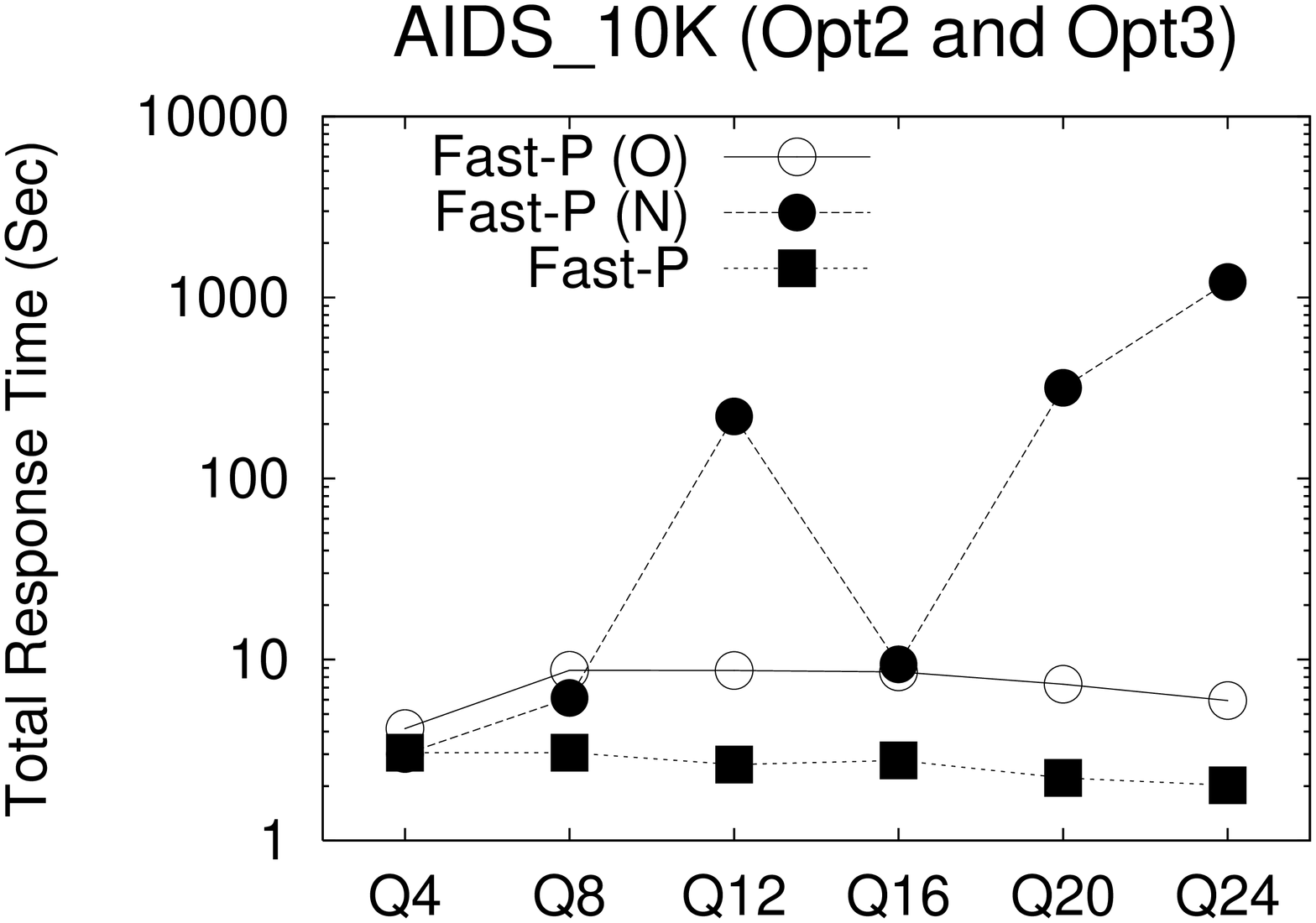,width=1.9in,height=4in,angle=-90}}
\caption{Effects of  Optimization in {\tt Fast-P} Algorithm}
\label{fig:effectopt1fast-p}
\end{figure}
To show the effect of the first optimization (Opt1),  we implemented
two versions, namely, {\tt Fast-P(1-Edge)} that sets $maxL = 1$ and
{\tt Fast-P(2-Edge)} that sets $maxL = 2$. Also, we use {\tt
Fast-ON} algorithm and denote it here by {\tt Fast-P(Vertex)} since
in  {\tt Fast-ON}, we apply
vertex-at-a-time-manner rather than path-at-a-time-manner. Figure
\ref{fig:effectopt1fast-p}(a) plots the results obtained by running
{\tt Fast-P(2-Edge)}, {\tt Fast-P(1-Edge)}, and {\tt Fast-P(Vertex)}
on AIDS\_10K for the different query sets. This figure  shows that
{\tt Fast-P(1-Edge)} is faster than {\tt Fast-P(Vertex)} except for
Q4, where {\tt Fast-P(Vertex)} shows the best performance. {\tt
Fast-P(2-Edge)} shows the best performance, it outperforms both {\tt
Fast-P(Vertex)} and {\tt Fast-P(1-Edge)}. This result is realistic
since {\tt Fast-P(2-Edge)} uses large-size local matches. Note that
{\tt Fast-P(Vertex)}, {\tt Fast-P(1-Edge)}, and {\tt Fast-P(2-Edge)}
apply the remaining two optimizations (Opt2 and Opt3). In the
following experiments, we denote {\tt Fast-P(2-Edge)} by {\tt
Fast-P}.\\

To show the effect of the remaining two optimizations (Opt2 and
Opt3), we implemented two versions, namely, {\tt Fast-P(N)}, that
uses the second optimization (Opt2) only and  {\tt Fast-P(O)}, that
uses the third optimization (Opt3) only. Note that, we set $maxL =
2$ for the two versions (i.e., the two versions apply the first
optimization). Figure \ref{fig:effectopt1fast-p}(b) plots the
results obtained by running the two versions on AIDS\_10K for the
different query sets. This figure shows that {\tt Fast-P(O)} is
faster than {\tt Fast-P(N)} except for Q4 and Q8, where {\tt
Fast-P(N)} shows the best performance. Note that, the third
optimization (Opt3) makes the algorithm ({\tt Fast-P(O)}) less
sensitive to query size. {\tt Fast-P} shows the best performance, it
outperforms both {\tt Fast-P(N)} and {\tt Fast-P(O)}.\\

The previous results confirm the fact that the three optimizations
in {\tt Fast-P} are neither independent nor conflicting, but they
are complementary to each other.
\end{itemize}
\begin{figure}[h]
\centering
\epsfig{file=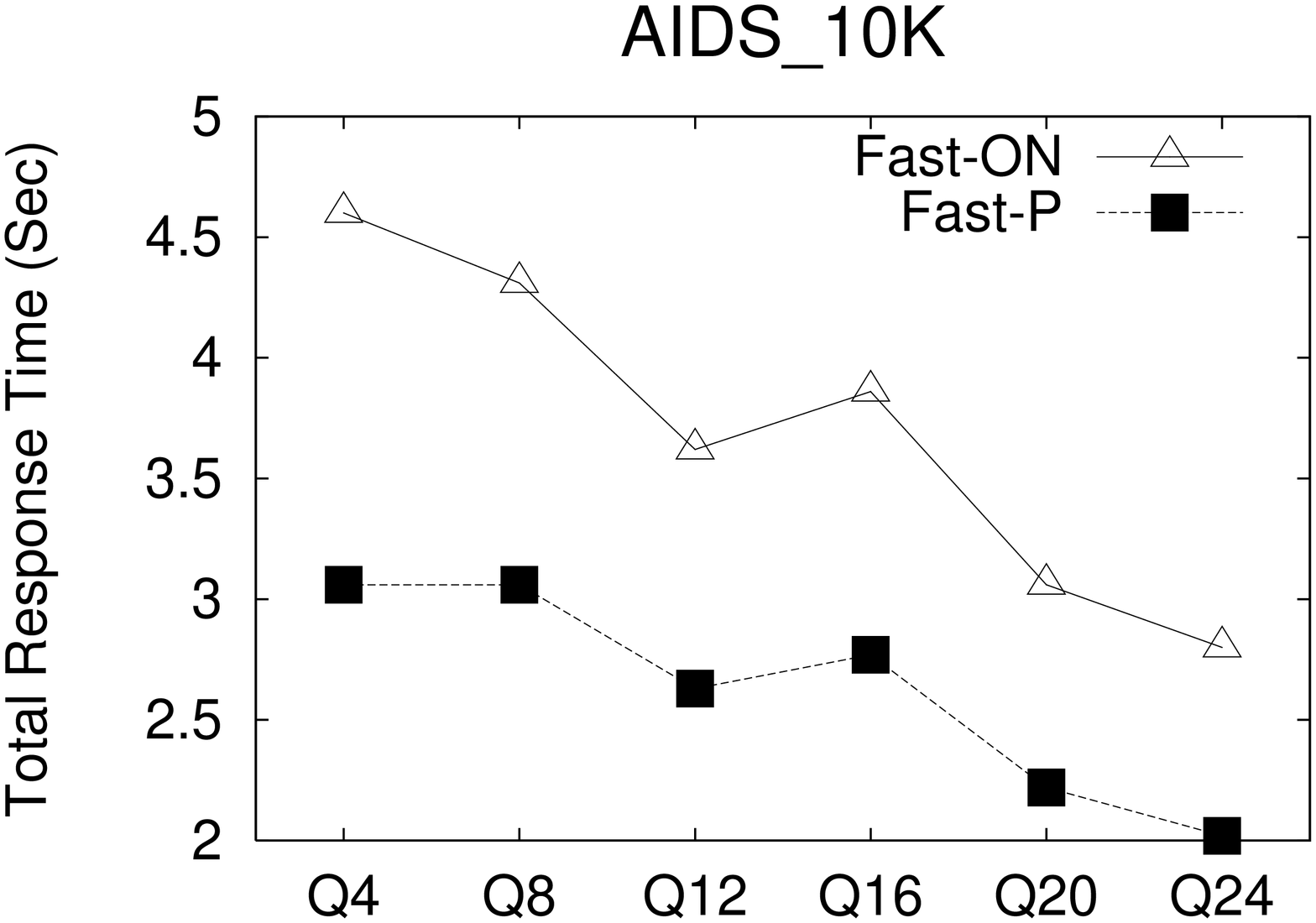,width=1.9in,height=4in,angle=-90}
\epsfig{file=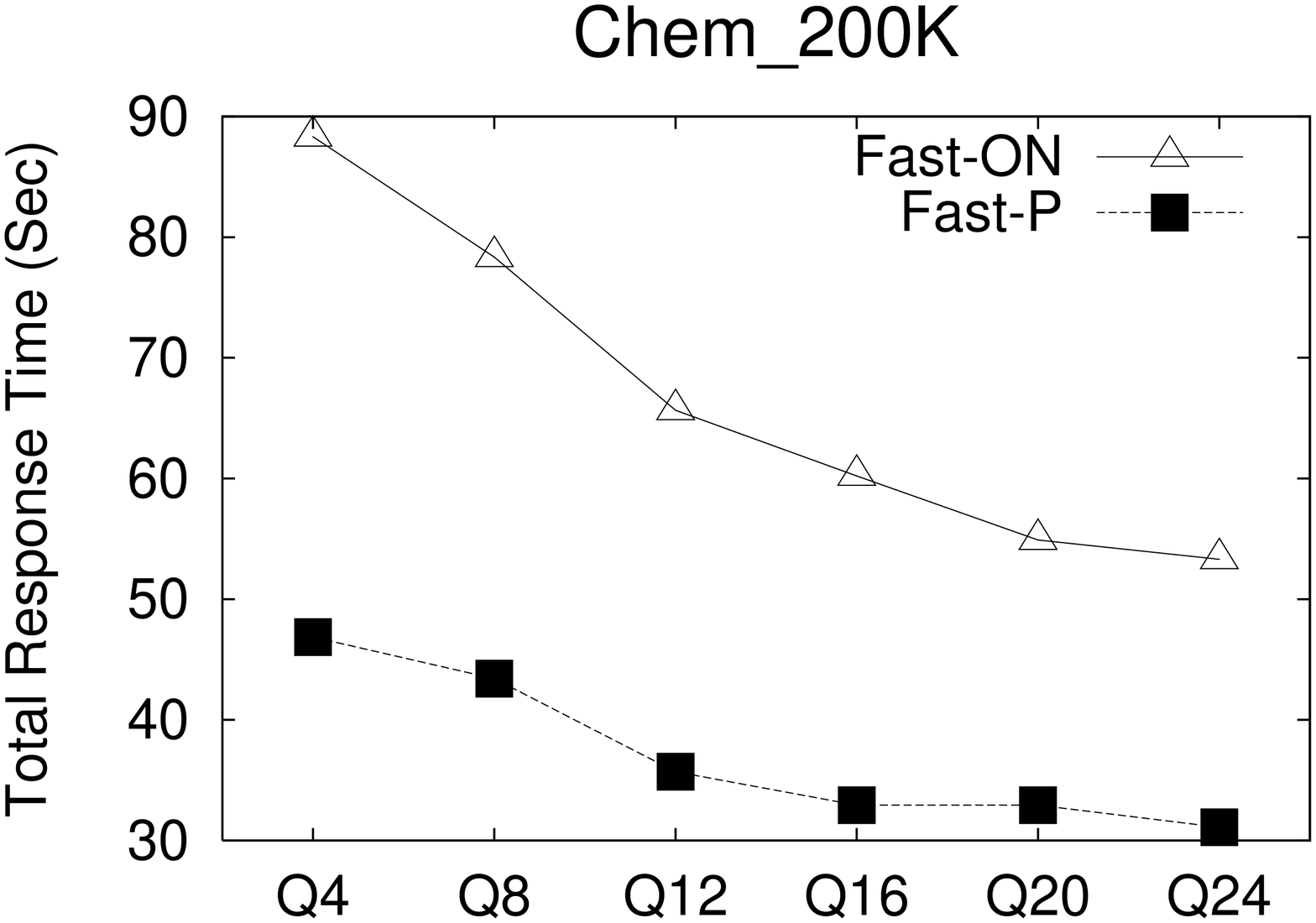,width=1.9in,height=4in,angle=-90}
\epsfig{file=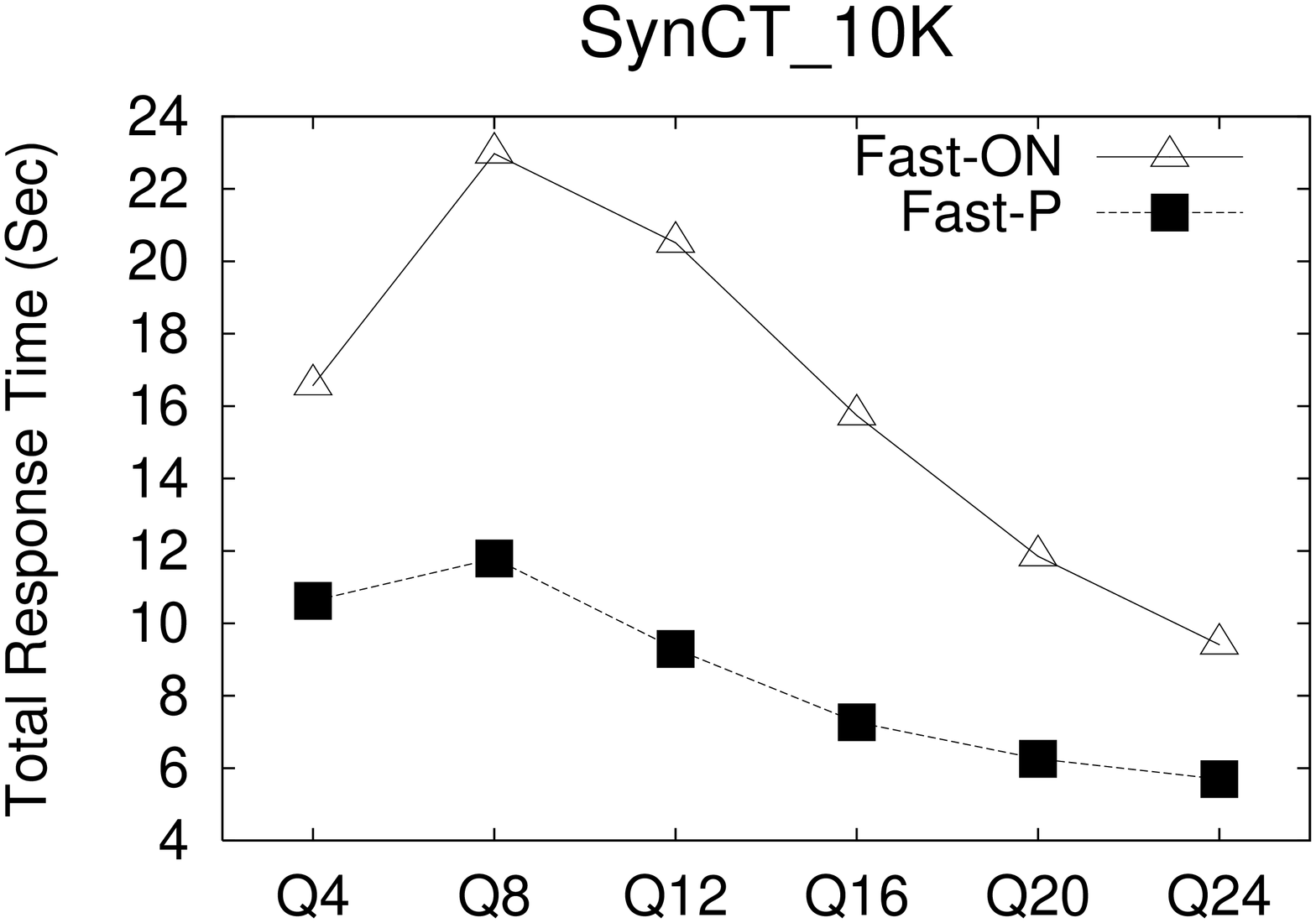,width=1.9in,height=4in,angle=-90}
\caption{Performance on Sparse Datasets ({\tt Fast-ON} vs. {\tt Fast-P})}
\label{fig:Fast-ON.vs.Fast-P_L}
\end{figure}
\begin{figure}
\epsfig{file=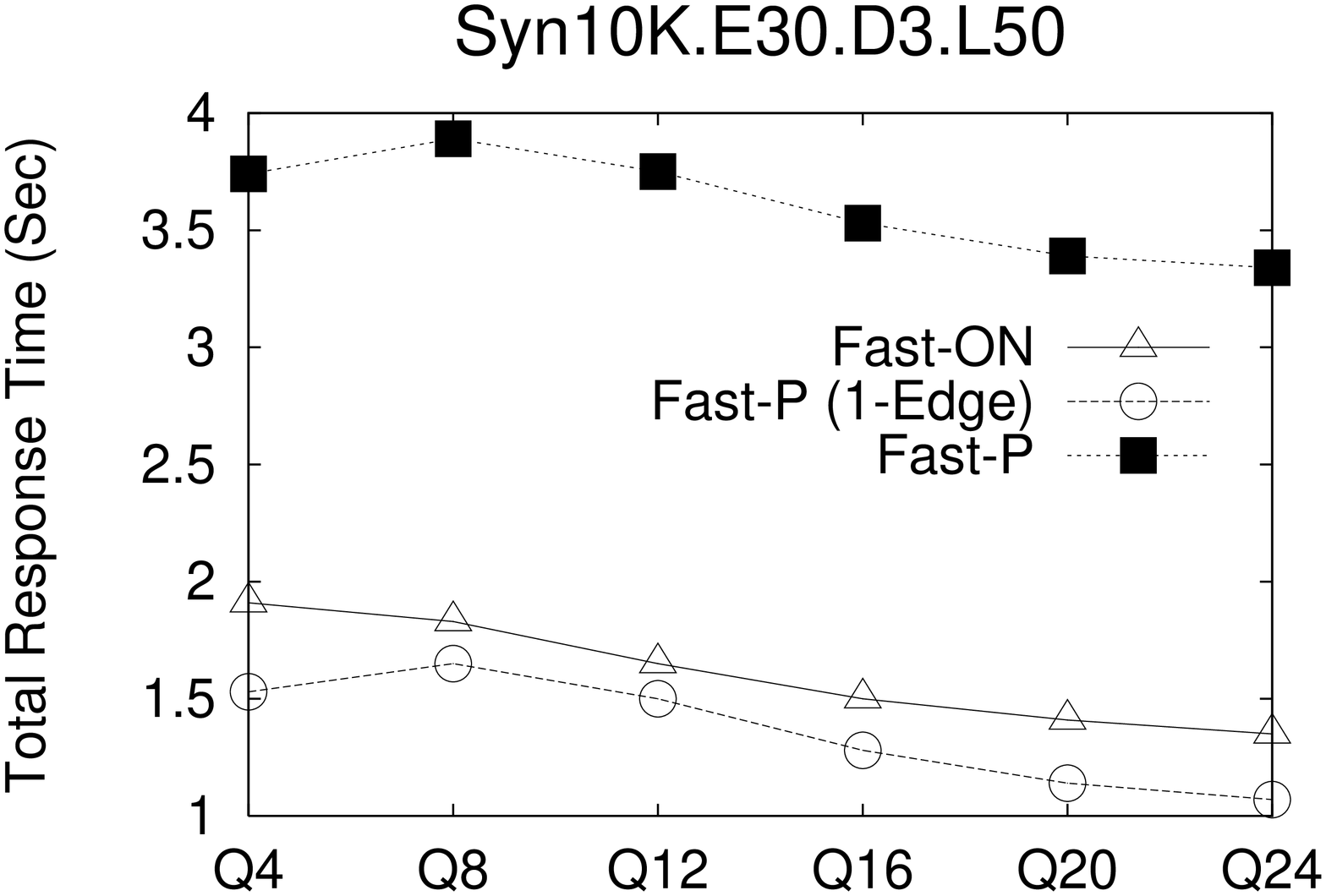,width=2.3in,height=3.1in,angle=-90}
\epsfig{file=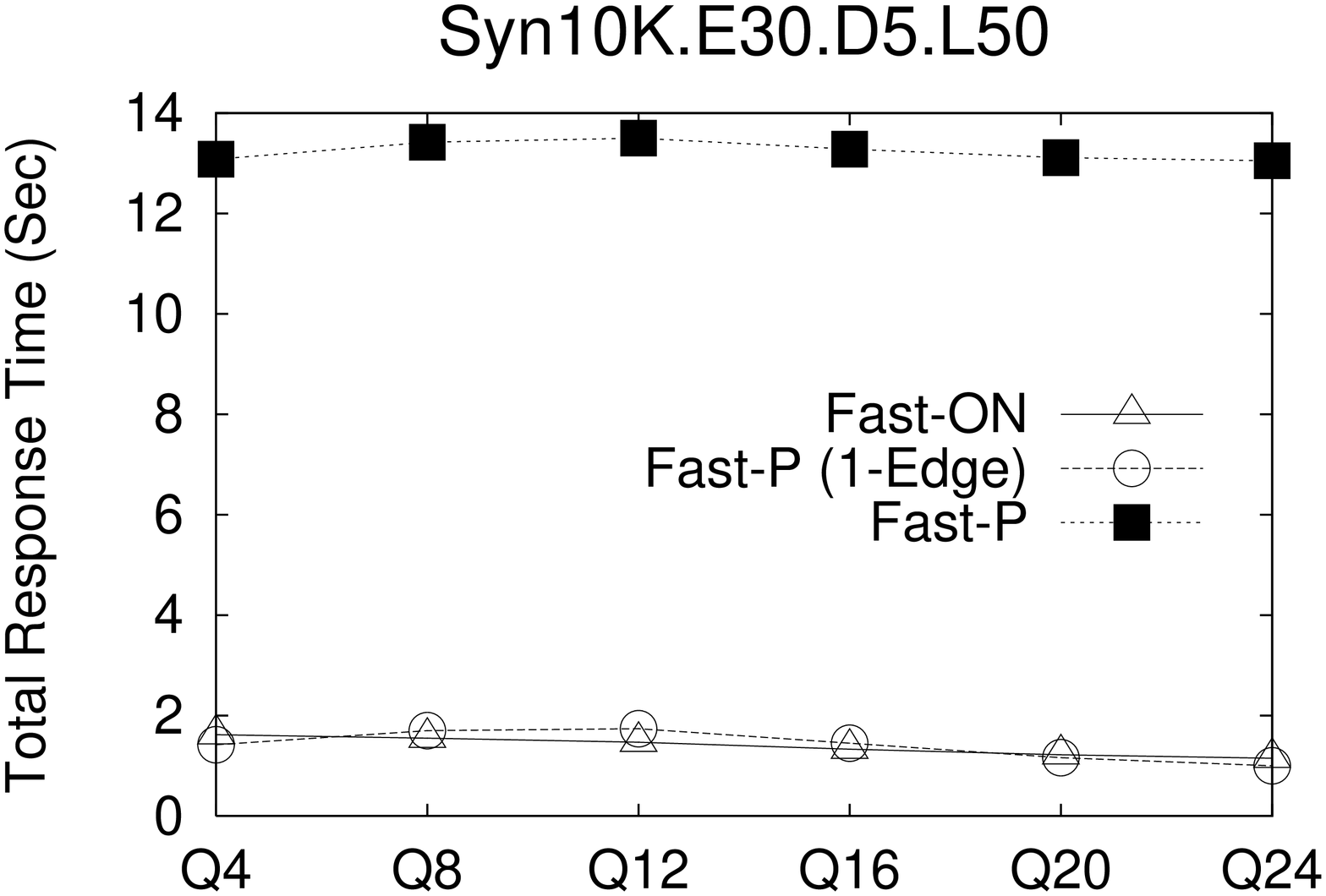,width=2.3in,height=3.1in,angle=-90}
\epsfig{file=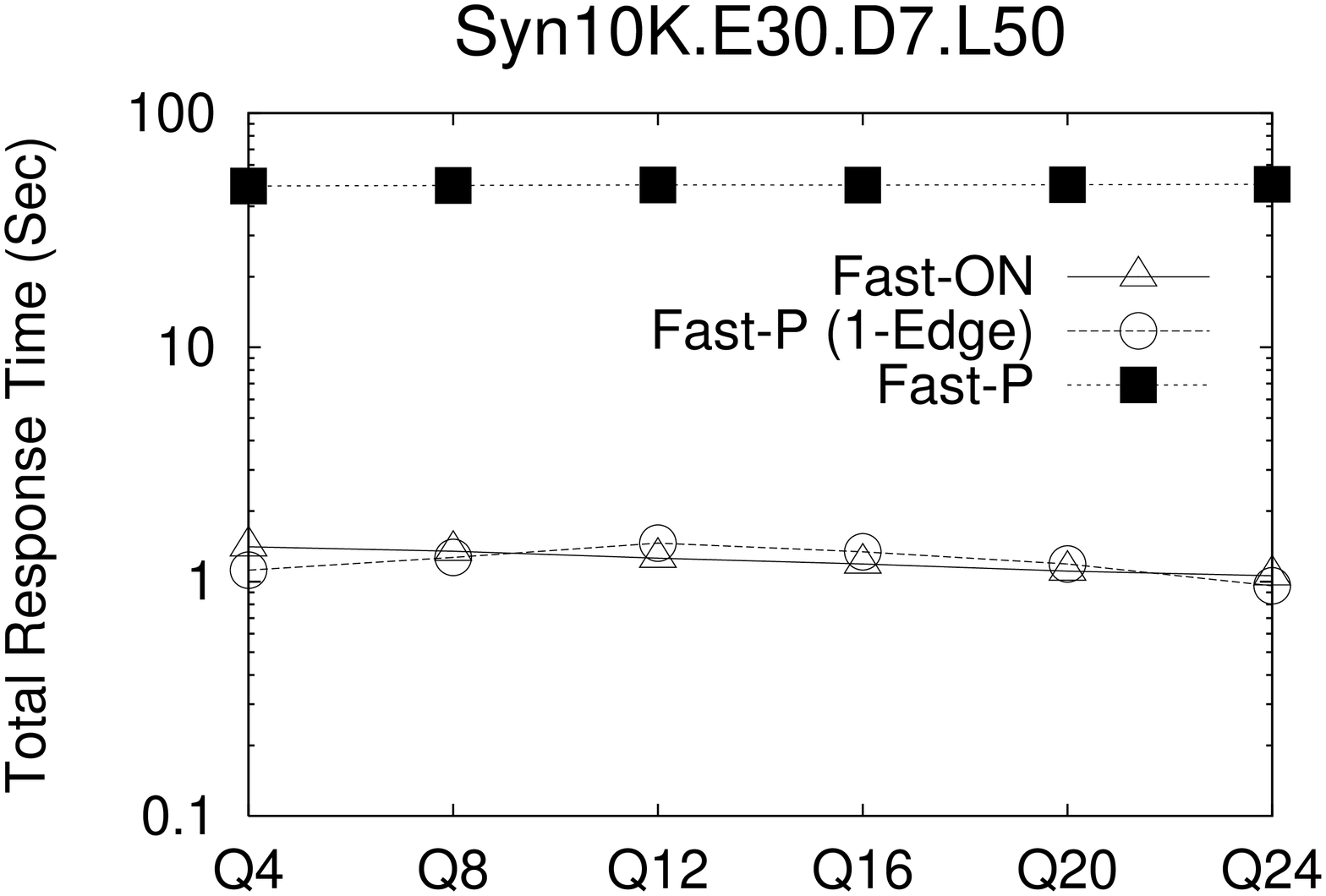,width=2.3in,height=3.1in,angle=-90}
\epsfig{file=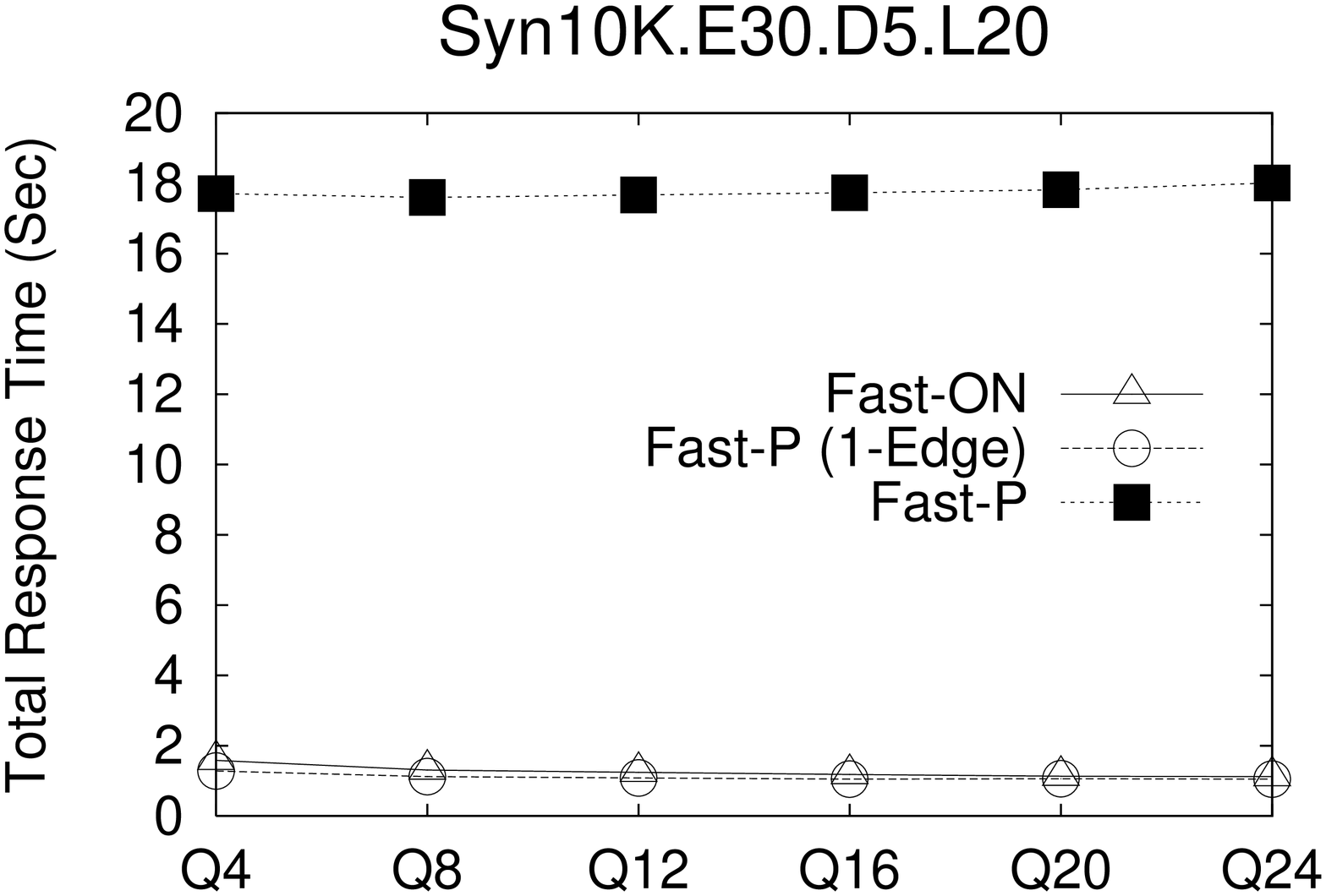,width=2.3in,height=3.1in,angle=-90}
\begin{center}
\epsfig{file=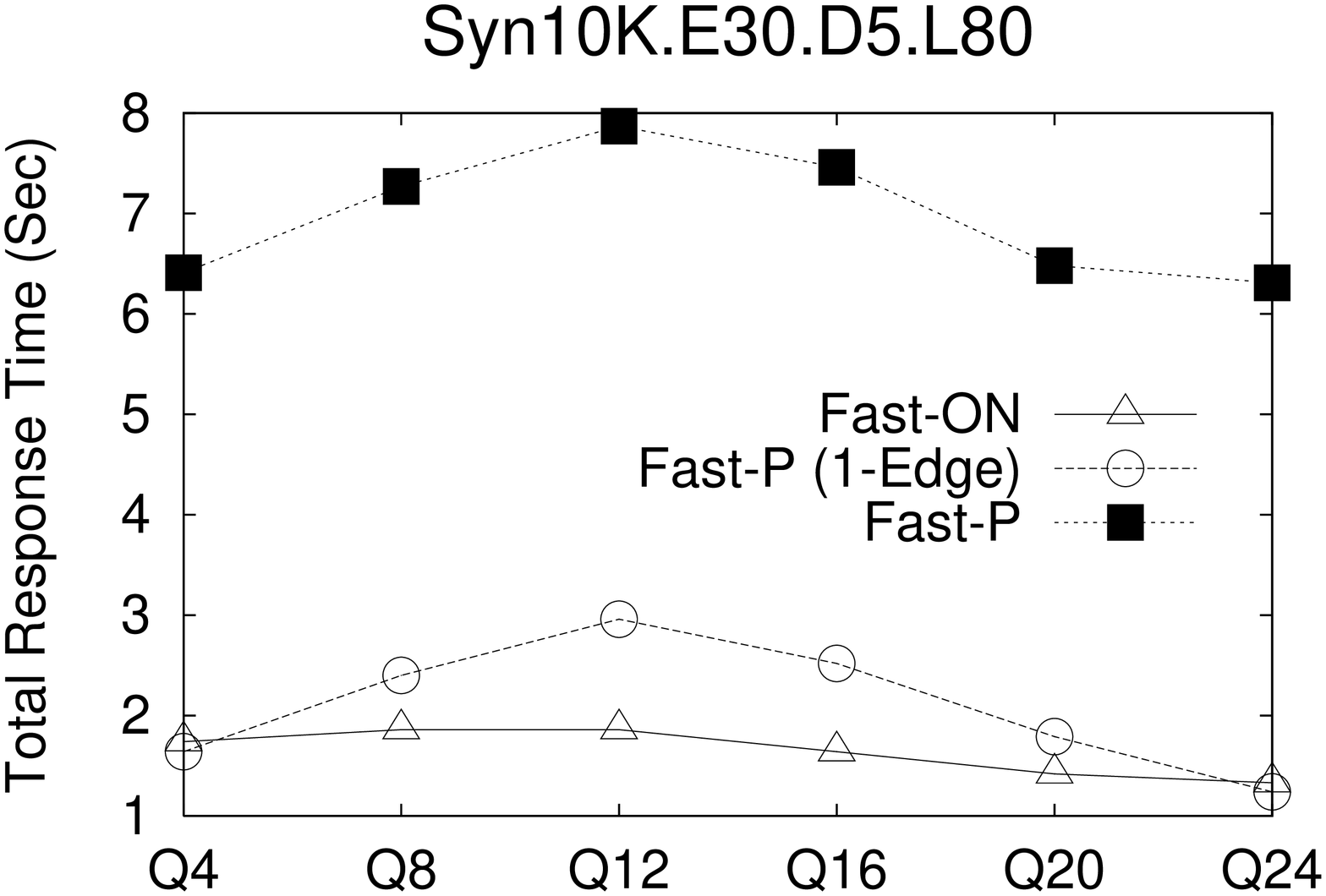,width=2.3in,height=3.1in,angle=-90}
\end{center}
\caption{Performance on Dense Datasets ({\tt Fast-ON} vs. {\tt Fast-P})} 
\label{fig:Fast-ON.vs.Fast-P1}
\end{figure}
\begin{figure}
\begin{center}
\epsfig{file=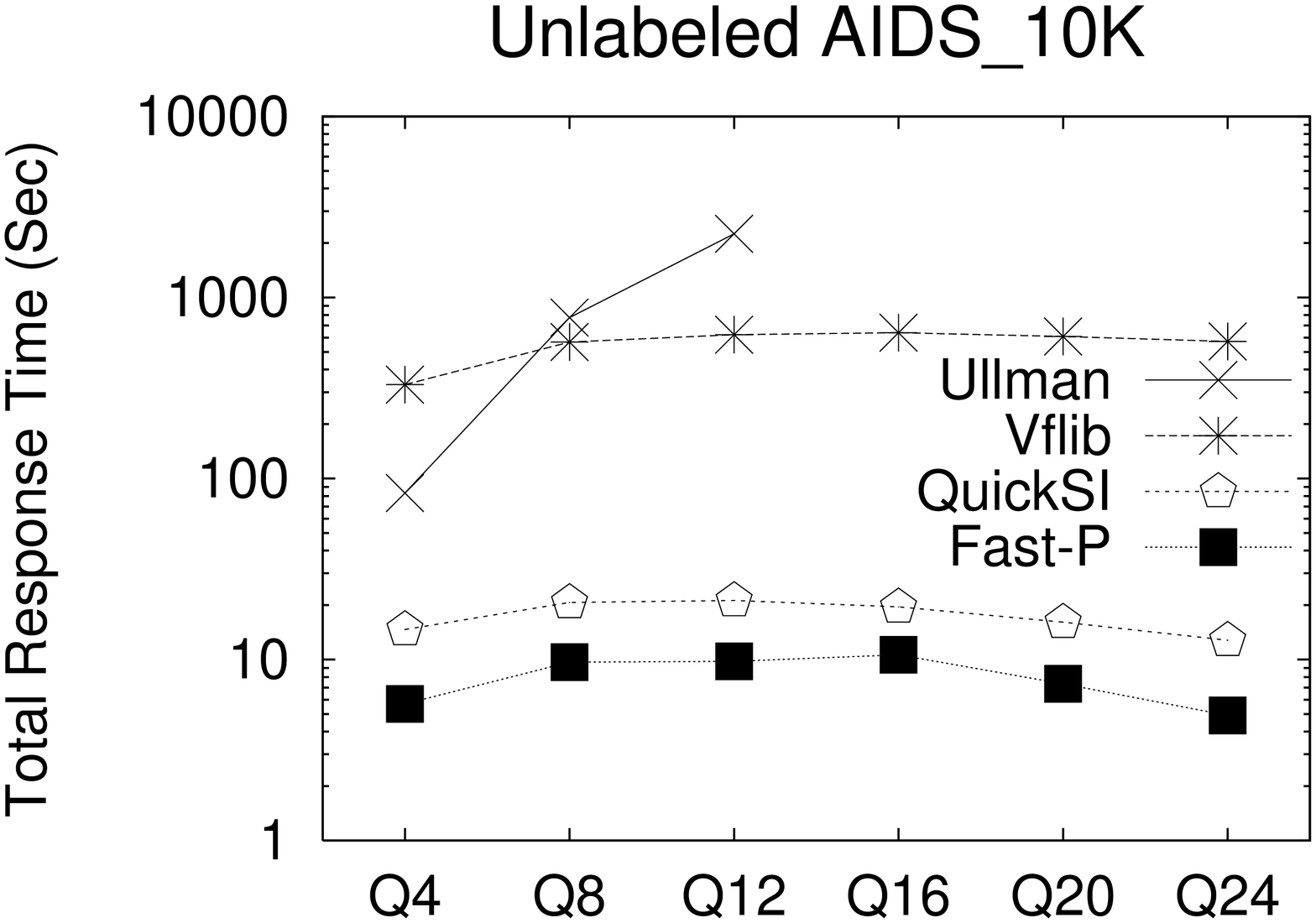,width=1.9in,height=4in,angle=-90}
\epsfig{file=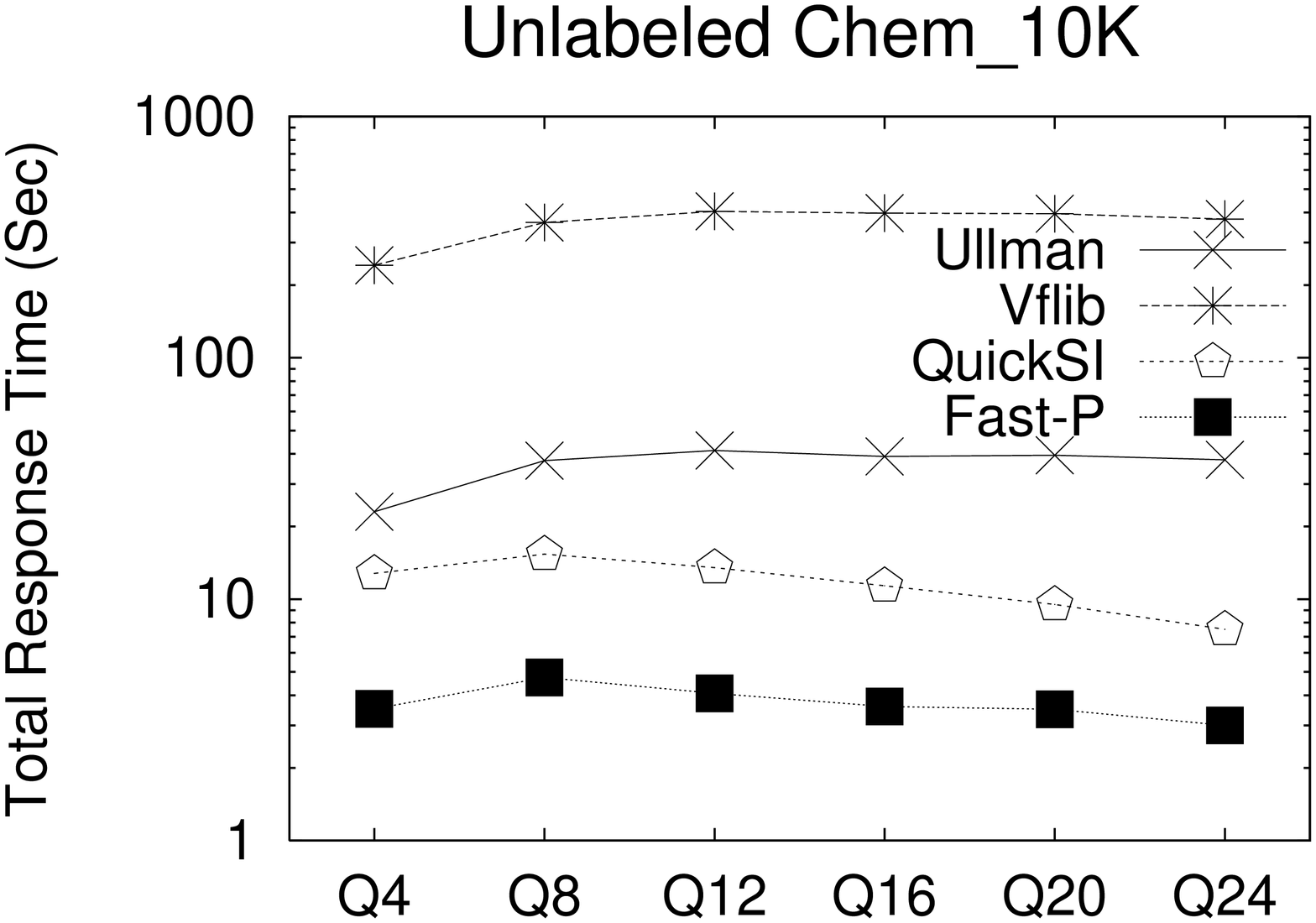,width=1.9in,height=4in,angle=-90}
\end{center}
\caption{Performance of {\tt Fast-P} on Sparse Unlabeled Datasets}
\label{fig:Fast-P_Sparse_Unlabeled_Datasets}
\end{figure}
\subsubsection{{\tt Fast-ON} vs. {\tt Fast-P}} In this section, we
demonstrate the efficiency of our two subgraph isomorphism
algorithms {\tt Fast-ON} and {\tt Fast-P} on sparse datasets (the
graphs have small density) and on dense datasets (the graphs have
high density) as follows.
\begin{itemize}
\item {\emph{Performance on Sparse Datasets}}\\ In this experiment, we
test the performance of {\tt Fast-ON} and {\tt Fast-P} on the sparse
datasets AIDS\_10K, Chem\_200K, and SynCT\_10K. Figure
\ref{fig:Fast-ON.vs.Fast-P_L} reports the results on these datasets.
From this figure, the {\tt Fast-P} algorithm always spends less
response time compared with {\tt Fast-ON} algorithm with a factor up
to 2. In the following experiments, for sparse datasets, we will use
{\tt Fast-P}.
\item {\emph{Performance on Dense Datasets}}\\ In this experiment, we
test the performance of {\tt Fast-ON}, {\tt Fast-P(1-Edge)}, and
{\tt Fast-P} on the five dense datasets Syn10K.E30.D3.L50,
Syn10K.E30.D5.L50,\\ Syn10K.E30.D7.L50, Syn10K.E30.D5.L80 and
Syn10K.E30.D5.L20. Figure \ref{fig:Fast-ON.vs.Fast-P1} reports the
results on the  five datasets. From this figure, we found that {\tt
Fast-P} algorithm is the worst one since both {\tt Fast-ON} and {\tt
Fast-P(1-Edge)} significantly outperform {\tt Fast-P}
algorithm. Roughly, both {\tt Fast-ON} and {\tt Fast-P(1-Edge)} have
the same response time on the  five datasets. In the following
experiments, we will use {\tt Fast-ON} for dense datasets. Note that
the performance gain of {\tt Fast-ON} against {\tt Fast-P}
dramatically increases when the density increases. This result is
occurred for the following two reasons. The first one is  due the
cost of inclusion tests  in {\tt Fast-P} since we can not use the
distinct neighborhood strategy with dense datasets. The second
reason is the large number of compatible paths to each query path.\\
\end{itemize}
\begin{figure}[h]
\centering
\epsfig{file=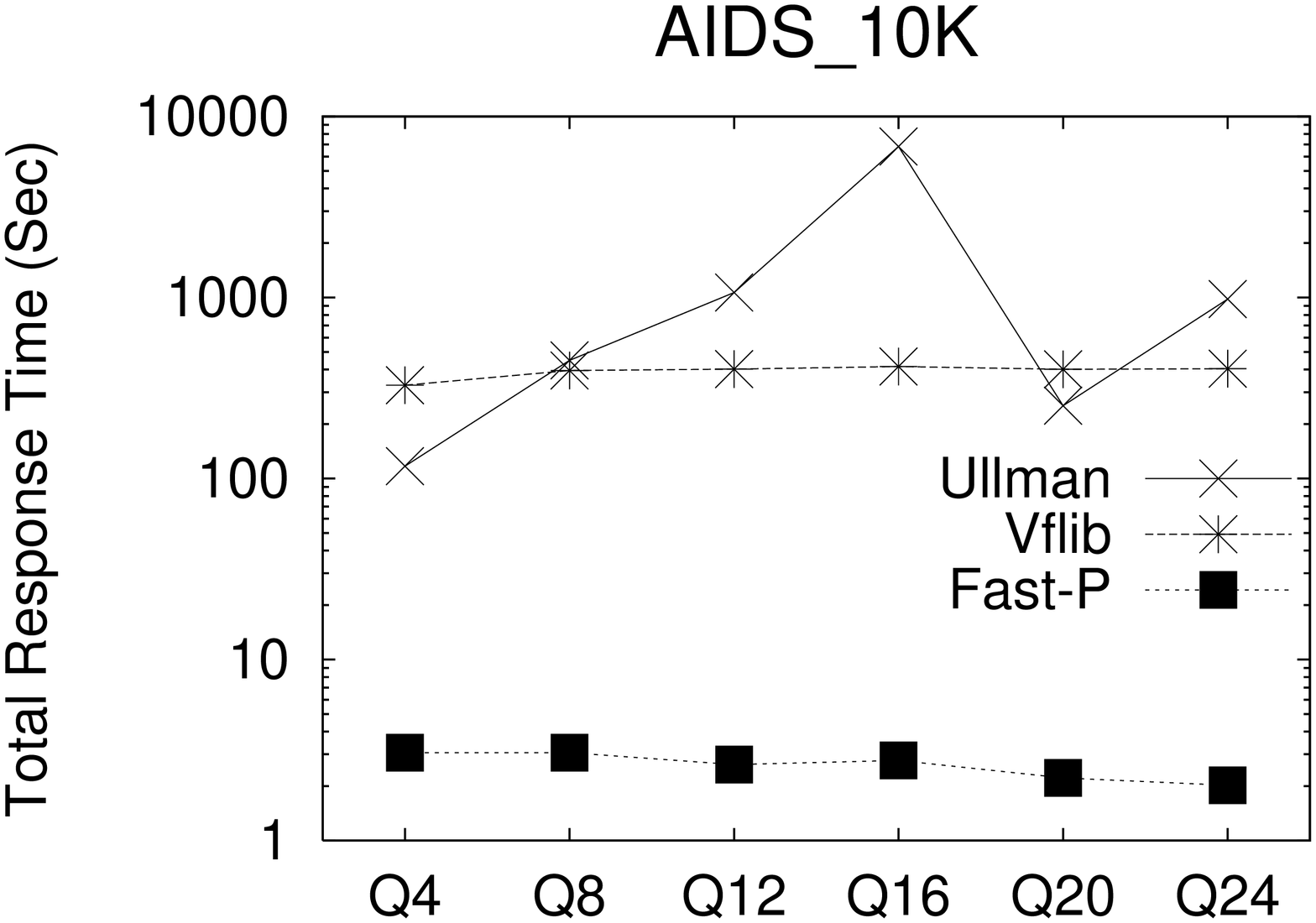,width=1.9in,height=4in,angle=-90}
\epsfig{file=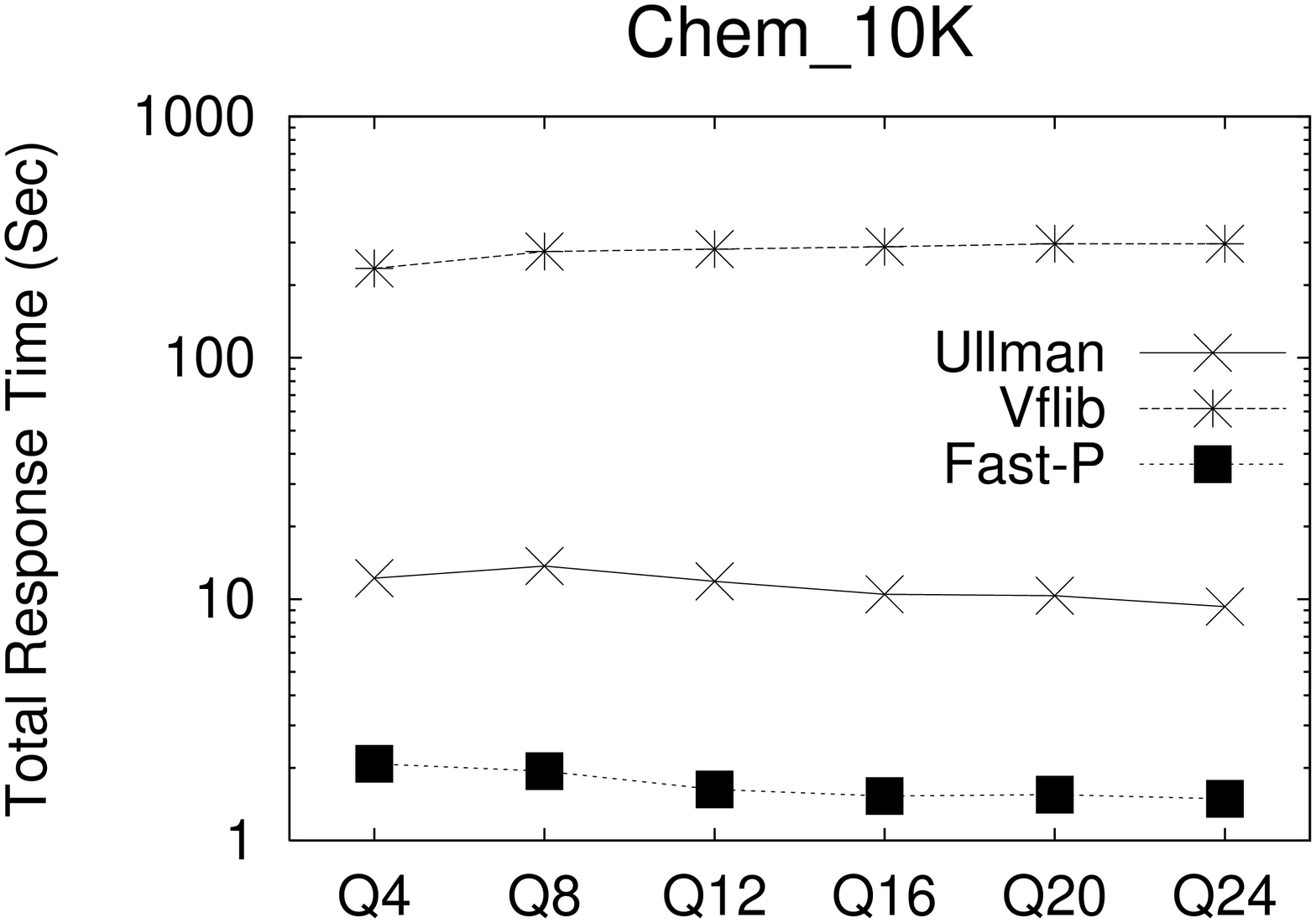,width=1.9in,height=4in,angle=-90}
\epsfig{file=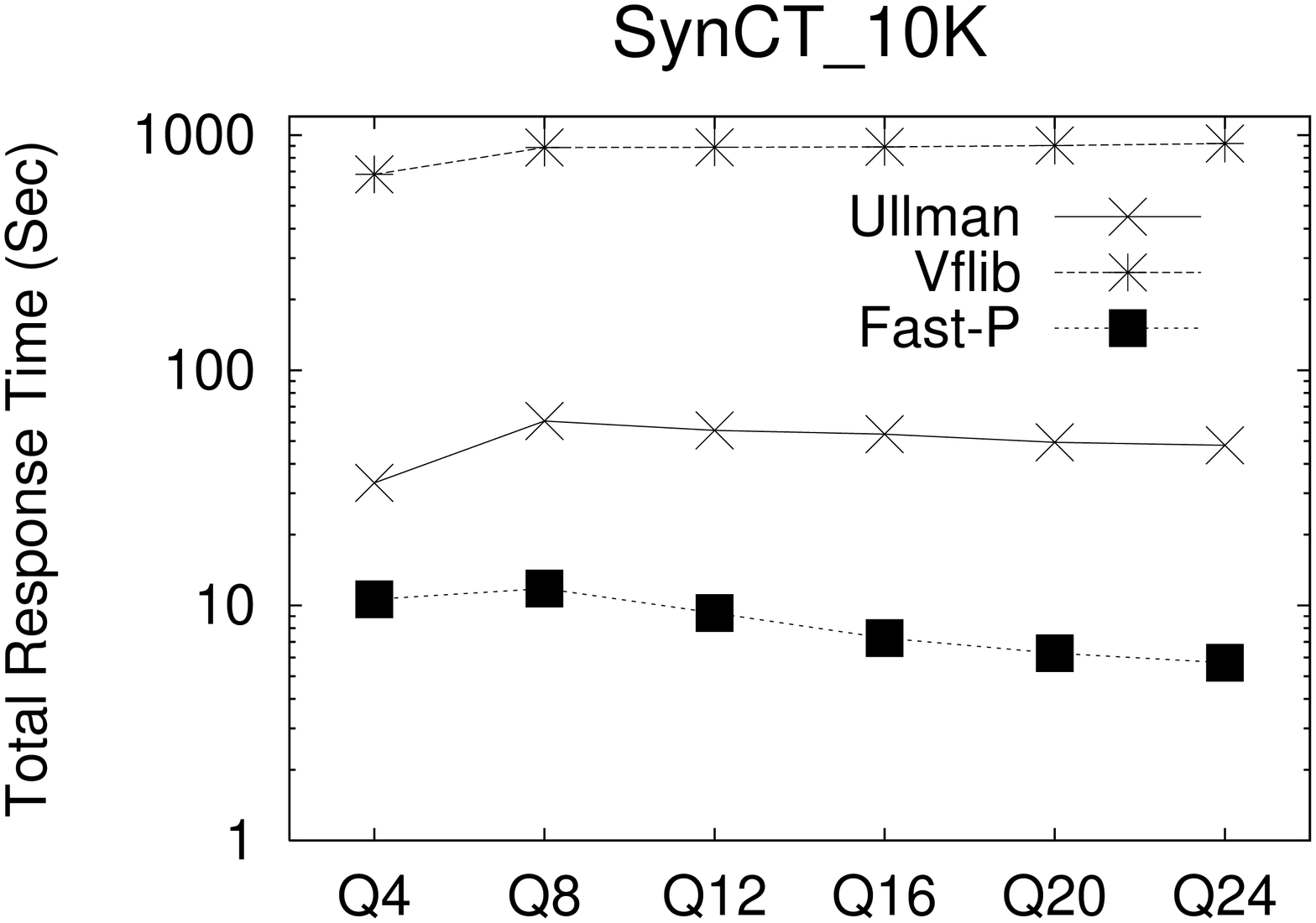,width=1.9in,height=4in,angle=-90}
\caption{Performance of {\tt Fast-P} on Sparse Labeled Datasets}
\label{fig:Fast-P_Sparse_Labeled_Datasets}
\end{figure}
In the next experiments, the two algorithms {\tt Fast-ON} and {\tt Fast-P} are tested against the
state-of-the-art subgraph isomorphism algorithms like Ullman (we implemented it using standard C++ with
STL library support),  QuickSI (we obtained its executable from the authors)
and Vflib (we downloaded it from http://amalfi.dis.unina.it/graph/db/vflib-2.0).

\subsubsection{{\tt Fast-P} vs. Ullman, Vflib, and QuickSI on Sparse
Datasets} In  this experiment, we demonstrate the efficiency of our
subgraph isomorphism testing algorithm {\tt Fast-P} against Ullman
and Vflib algorithms on labeled sparse datasets and against Ullman,
Vflib, and QuickSI (works with unlabeled edges datasets only) algorithms on unlabeled sparse datasets as
follows.
\begin{itemize}
\item {\emph{On Labeled Sparse Datasets}}\\ Here, we
evaluate the performance of {\tt Fast-P} on AIDS\_10K, Chem\_10K,
and SynCT\_10K datasets by comparing it with the two algorithms
Ullman and Vflib. Total response time  for each query set of the
three datasets is recorded in Figure
\ref{fig:Fast-P_Sparse_Labeled_Datasets}. For the two datasets
Chem\_10K and SynCT\_10K, Ullman is faster than Vflib while Vflib
outperforms Ullman on AIDS\_10K except for Q4. {\tt Fast-P} shows
the best performance, it outperforms both Ullman and Vflib on the
three dataset with a wide margin.
\item {\emph{On Unlabeled Sparse Datasets}}\\ Here, we
We used the two sparse datasets AIDS\_10K and Chem\_10K after
removing the edge labels and we denoted them as Unlabeled AIDS\_10K
and Unlabeled Chem\_10K. Figure
\ref{fig:Fast-P_Sparse_Unlabeled_Datasets} reports the results on
the two datasets. From this figure, QuickSI outperforms Ullman and
Vflib on the two datasets. Also, {\tt Fast-P} shows the best
performance, it outperforms Ullman, Vflib, and QuickSI on AIDS\_10K
dataset by more than two order of magnitude, more than one order of
magnitude, and three factors, respectively (Note that Ullman is not
shown for the query sets, namely, Q16, Q20, and Q24 since it failed
to run on our machine). On Chem\_10K dataset, {\tt Fast-P}
outperforms Ullman, Vflib, and QuickSI by one  order of magnitude,
more than two order of magnitud, and 4 factors, respectively.
\end{itemize}
\subsubsection{{\tt Fast-ON} vs. Ullman, Vflib, and QuickSI on Dense
Datasets} 
\begin{figure}
\epsfig{file=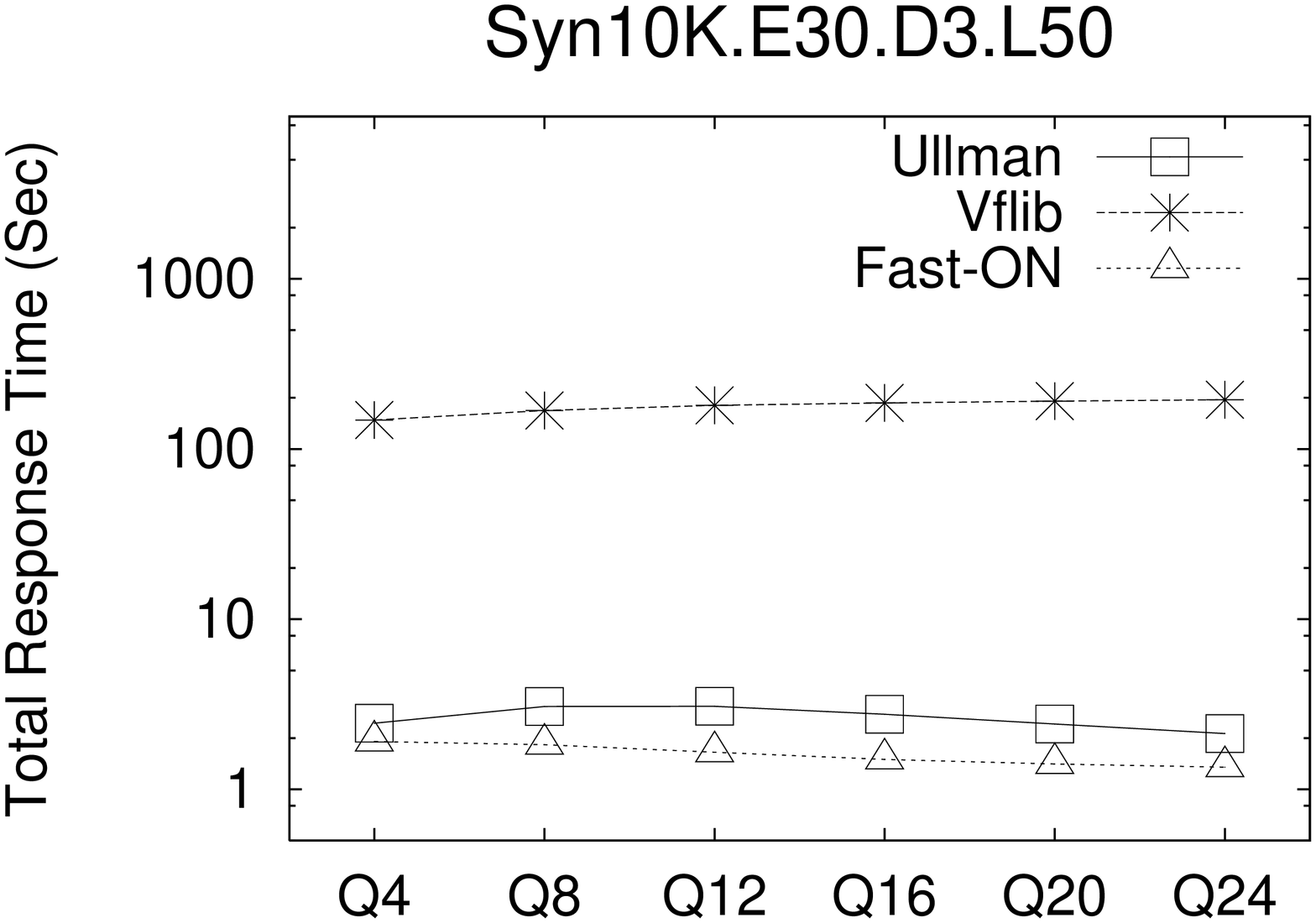,width=2.3in,height=3.1in,angle=-90}
\epsfig{file=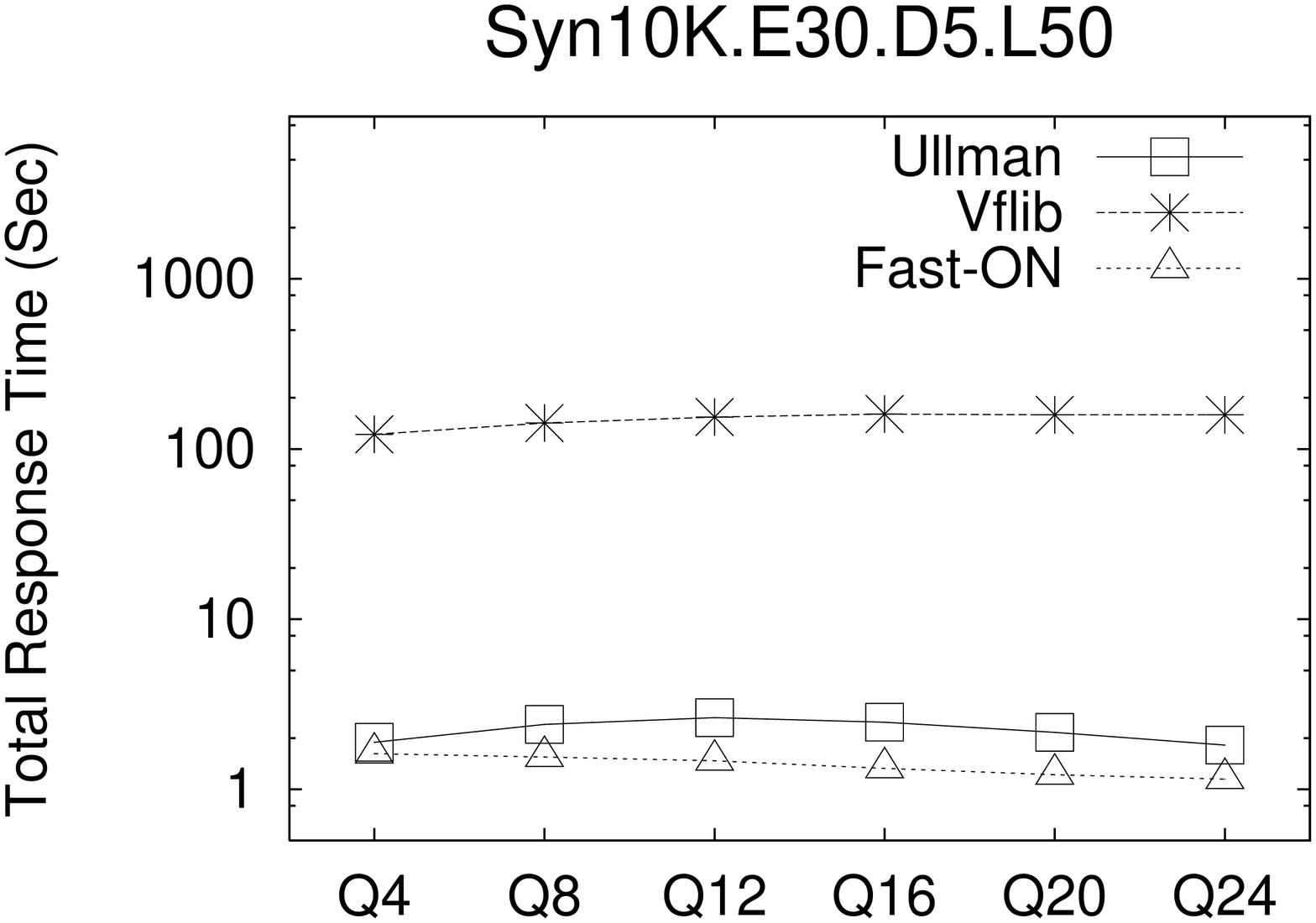,width=2.3in,height=3.1in,angle=-90}
\epsfig{file=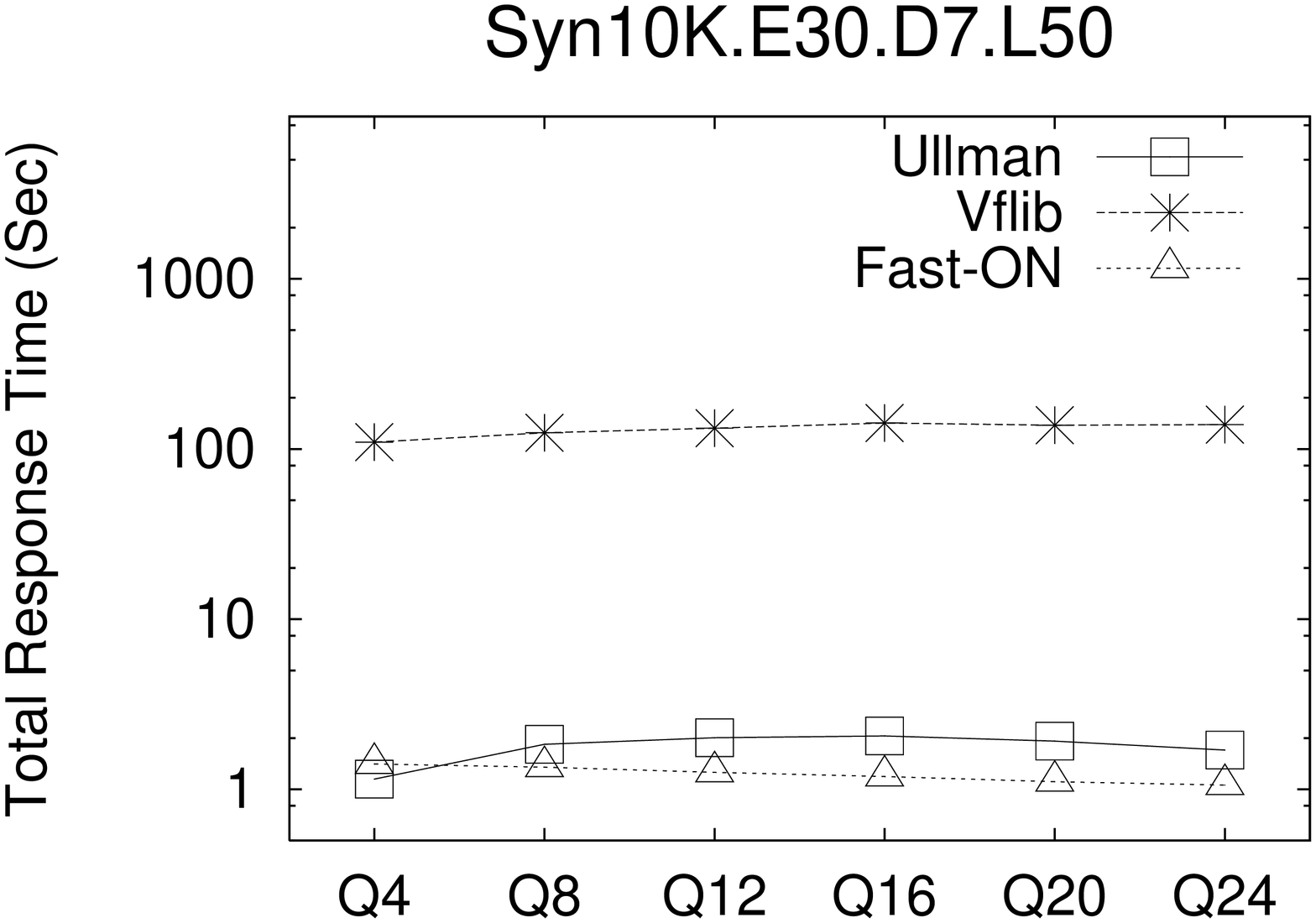,width=2.3in,height=3.1in,angle=-90}
\epsfig{file=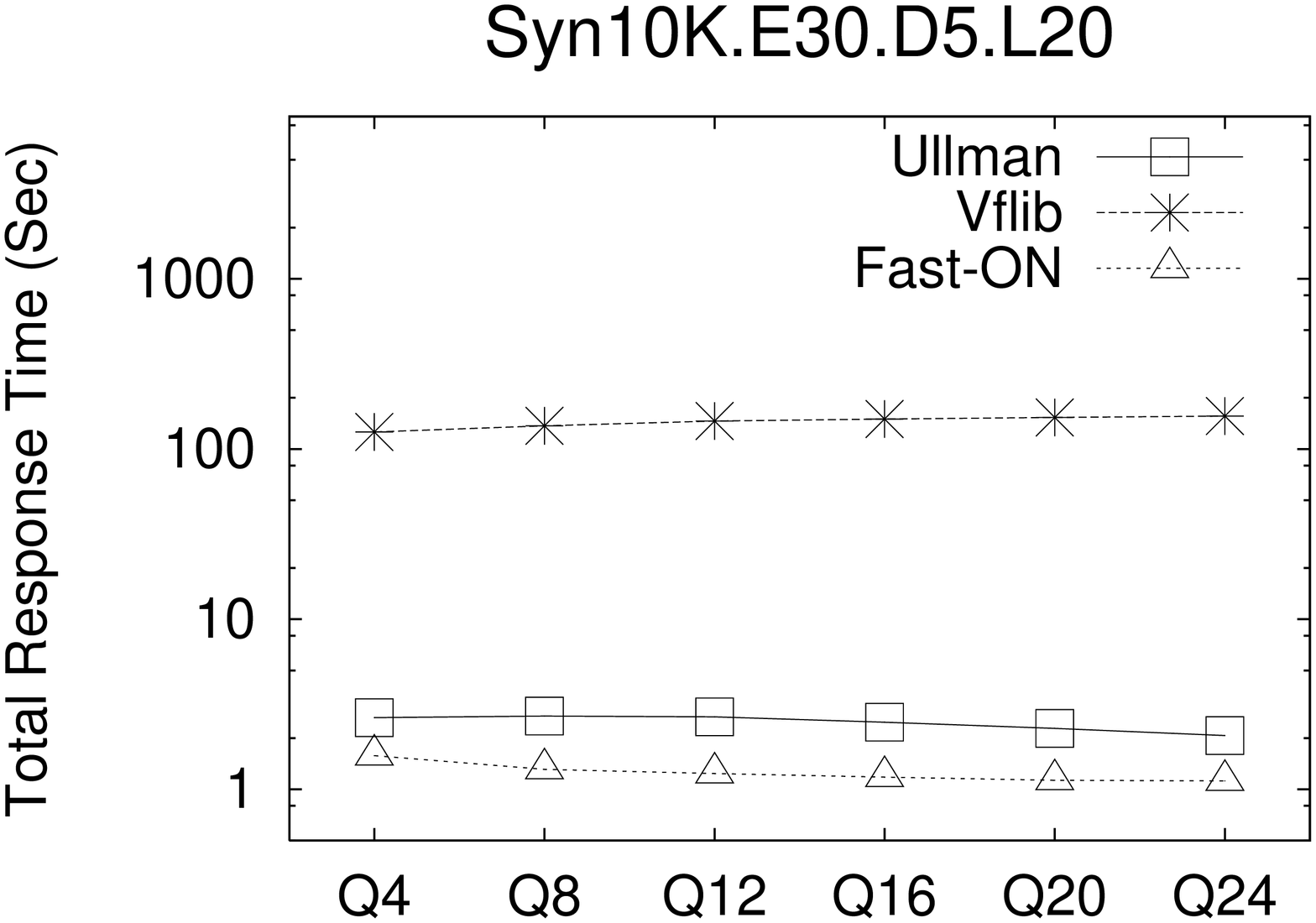,width=2.3in,height=3.1in,angle=-90}
\begin{center}
\epsfig{file=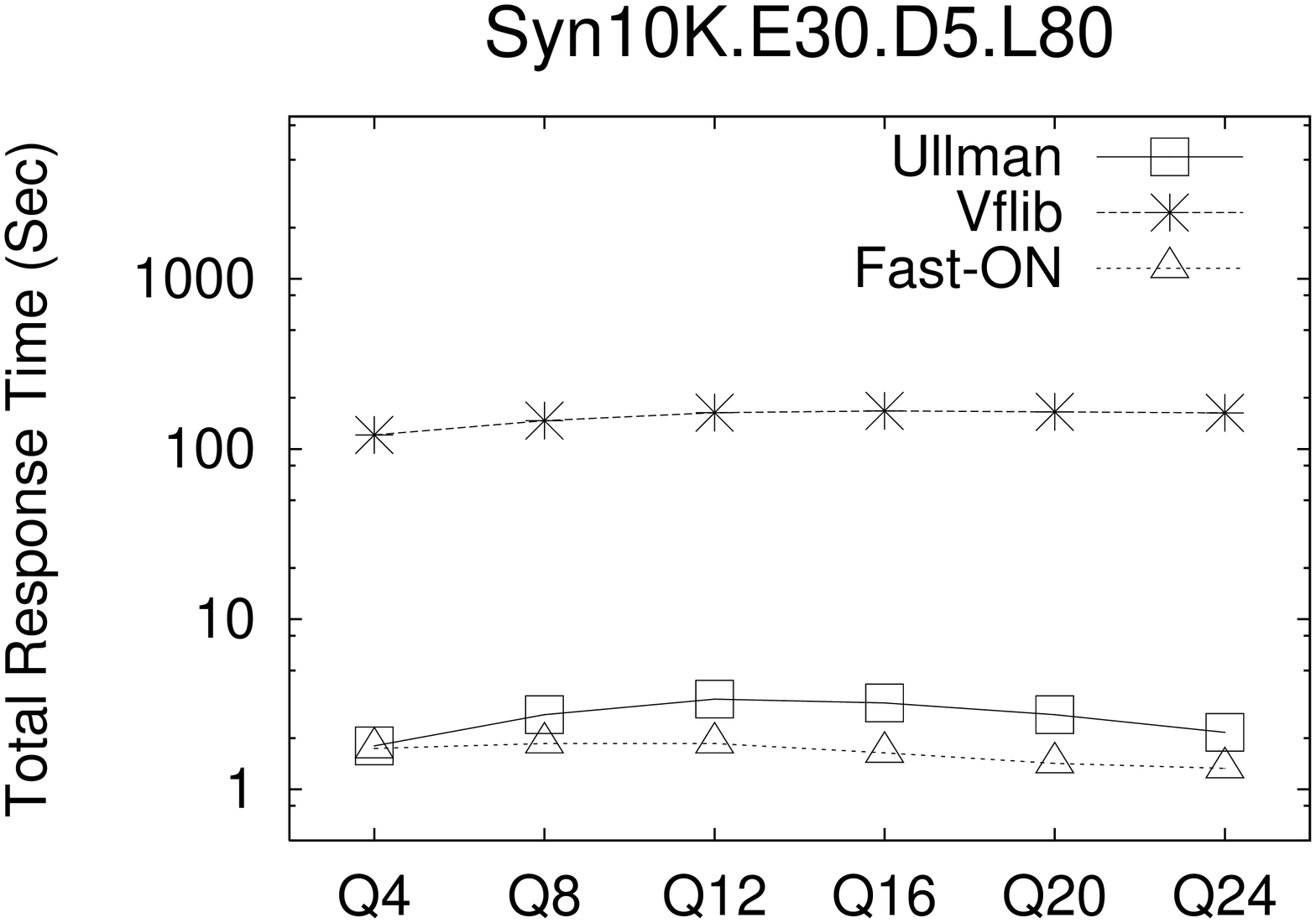,width=2.3in,height=3.1in,angle=-90}
\end{center}
\caption{Performance of {\tt Fast-ON} on Dense Labeled Datasets} 
\label{fig:Fast-ON-Performance_Dense_Datasets_Labeled} 
\end{figure}

\begin{figure}
\centering
\epsfig{file=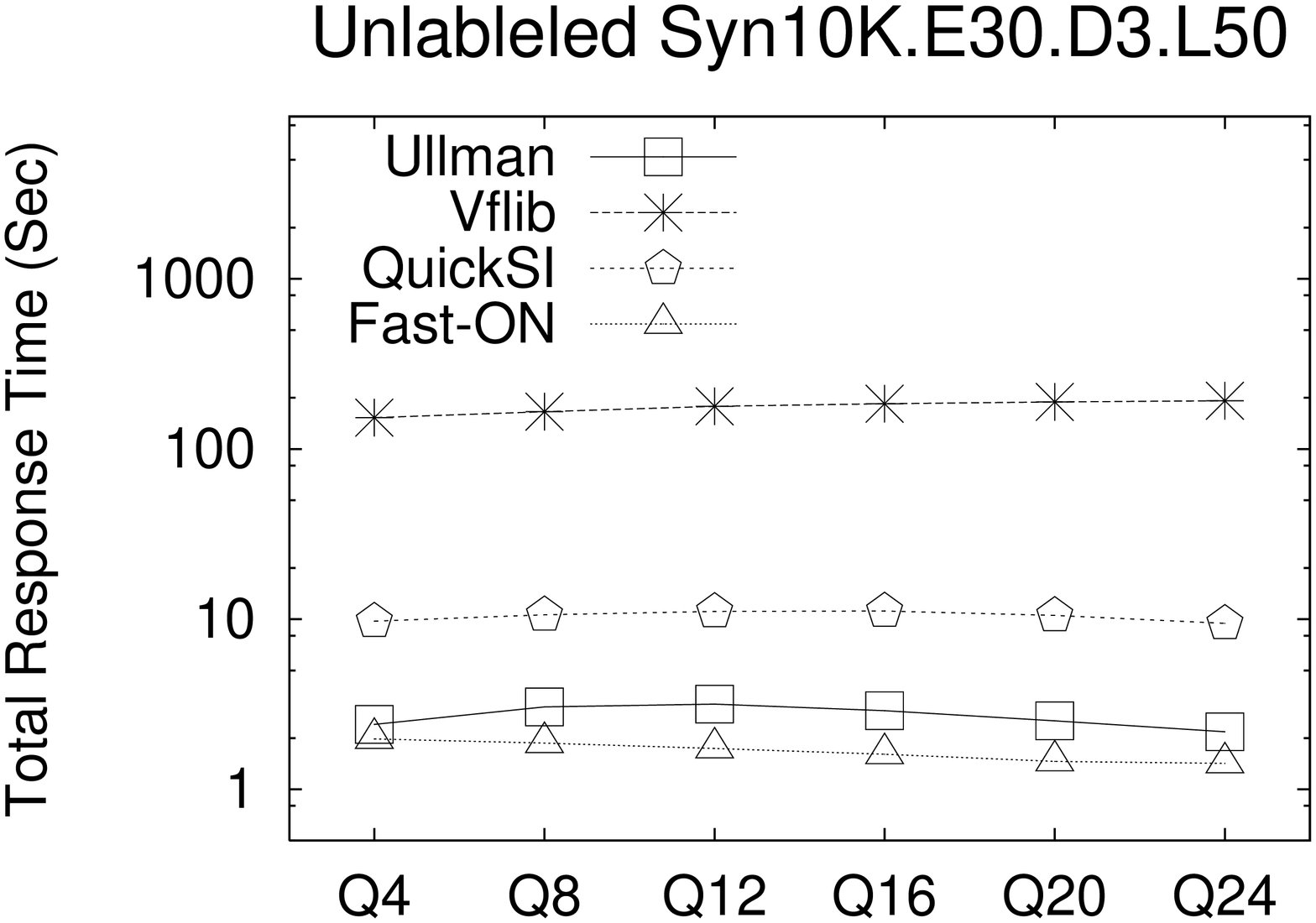,width=1.9in,height=4in,angle=-90}
\epsfig{file=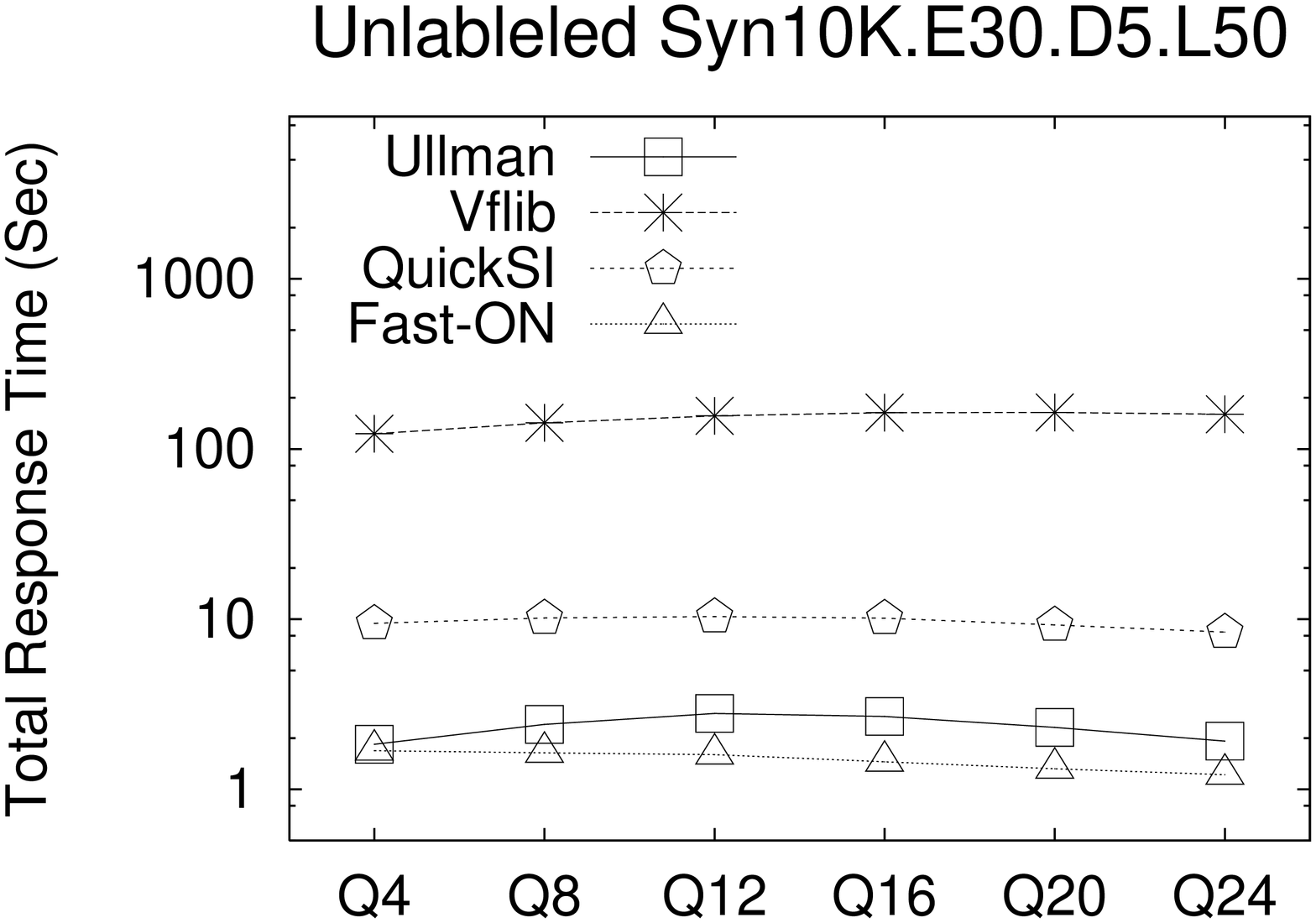,width=1.9in,height=4in,angle=-90}
\epsfig{file=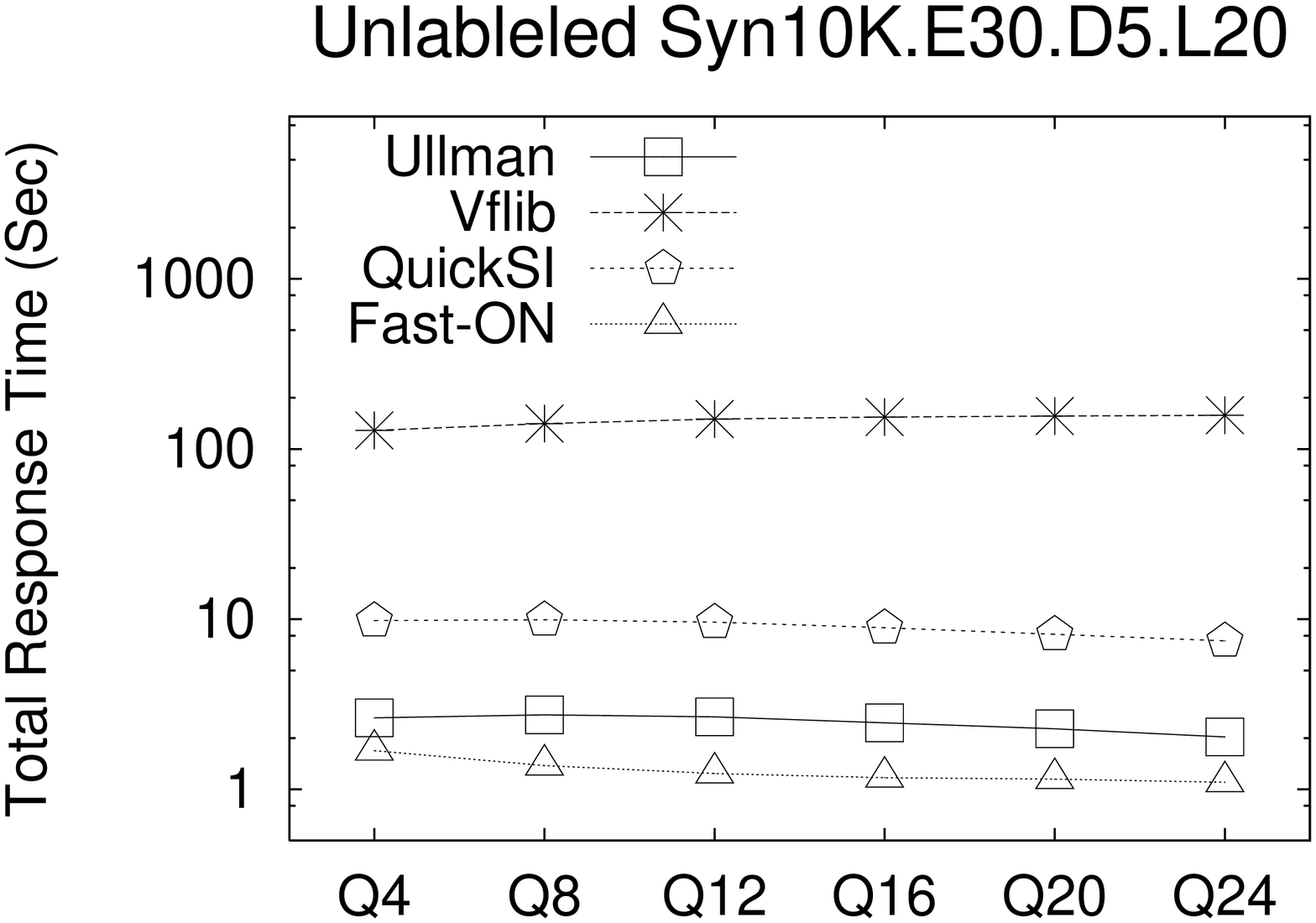,width=1.9in,height=4in,angle=-90}
\caption{Performance of {\tt Fast-ON} on Dense Unlabeled Datasets}
\label{fig:Fast-ON-Performance_Dense_Datasets_Unlabeled}
\end{figure}

In this experiment, we demonstrate the efficiency of our
subgraph isomorphism testing algorithm {\tt Fast-ON} against Ullman
and Vflib algorithms on labeled dense datasets and against Ullman,
Vflib, and QuickSI (works with unlabeled edges datasets only) algorithms on unlabeled dense datasets as
follows.
\begin{itemize}
\item {\emph{On Labeled Dense Datasets}}\\
In this subsection, we evaluate the performance of {\tt Fast-ON} on
the five dense datasets Syn10K.E30.D3.L50, Syn10K.E30.D5.L50,
Syn10K.E30.D7.L50, Syn10K.E30.D5.L80 and Syn10K.E30.D5.L20 by
comparing it with the two algorithms Ullman and Vflib. Total
response time  for each query set of the five datasets is recorded
and demonstrated in the Figure
\ref{fig:Fast-ON-Performance_Dense_Datasets_Labeled}. From this
figure, Ullman is faster than Vflib by a large margin and {\tt
Fast-ON} shows the best performance, it outperforms Ullman and Vflib
on the five labeled dense datasets by up to 3 factors and more than
two order of magnitude, respectively.
\item {\emph{On Unlabeled Dense Datasets}}\\
Here, we used the three dense datasets Syn10K.E30.D3.L50, Syn10K.E30.D5.L50,
and Syn10K.E30.D5.L20, after removing the edge labels and we denoted
them as Unlabeled Syn10K.E30.D3.L50, Unlabeled Syn10K.E30.D5.L50 and
Unlabeled Syn10K.E30.D5.L20. Total response time  for each query
set of the three datasets is recorded and demonstrated in the Figure
\ref{fig:Fast-ON-Performance_Dense_Datasets_Unlabeled}. From this
figure, Ullman outperforms Vflib and QuickSI on the three datasets,
Vflib is the worst one, and {\tt Fast-ON} shows the best
performance, it outperforms Ullman, Vflib, and QuickSI  on the three
unlabeled dense datasets by up to 3 factors, more than two order of
magnitude, and more than one order of magnitude, respectively.
\end{itemize}
\begin{figure}
\centering
\epsfig{file=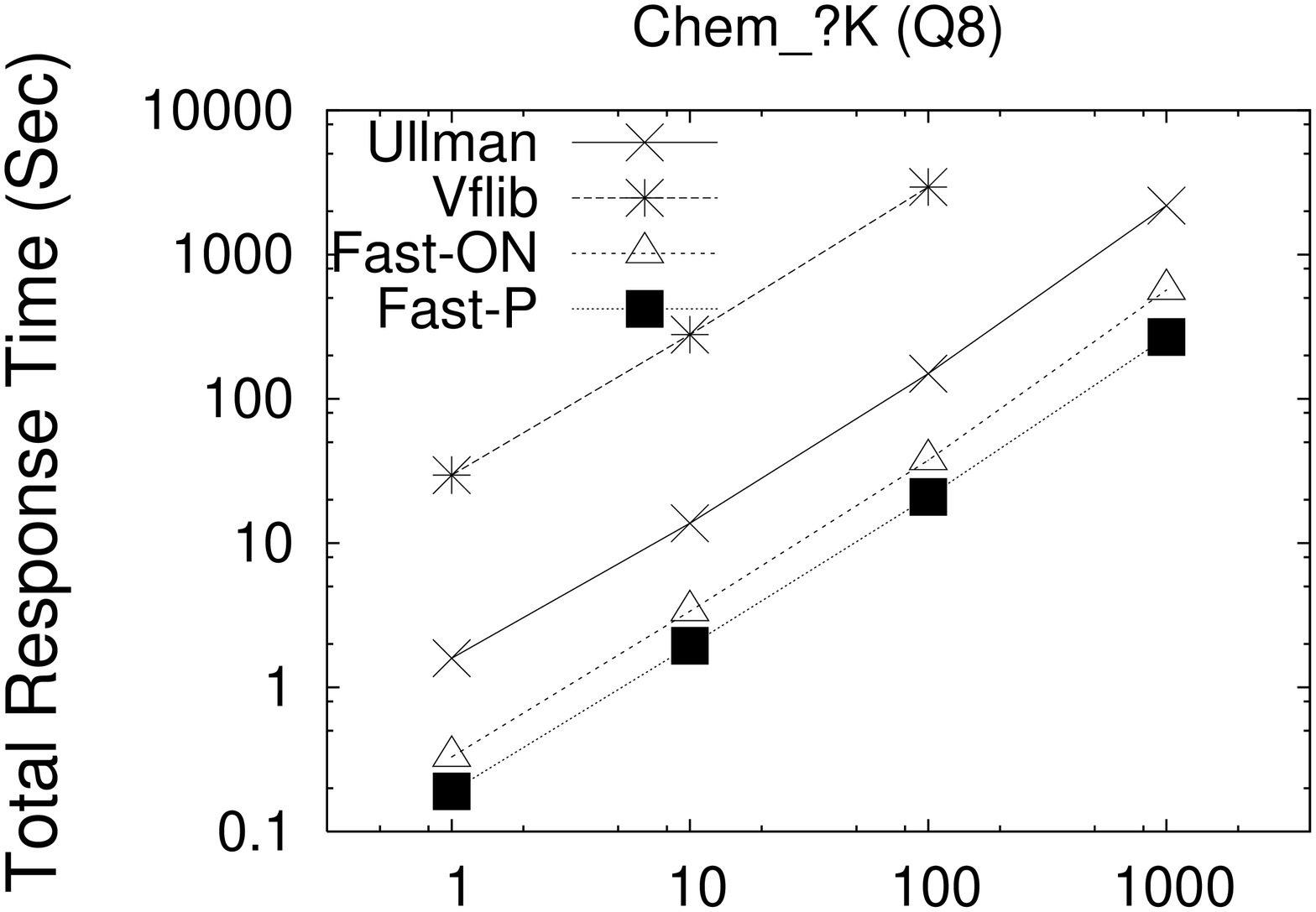,width=1.9in,height=4in,angle=-90}
\caption{Scalability on Dataset Size (\# Graphs in K)}
\label{fig:chemscalability}
\end{figure}
\begin{figure}
\centering
\epsfig{file=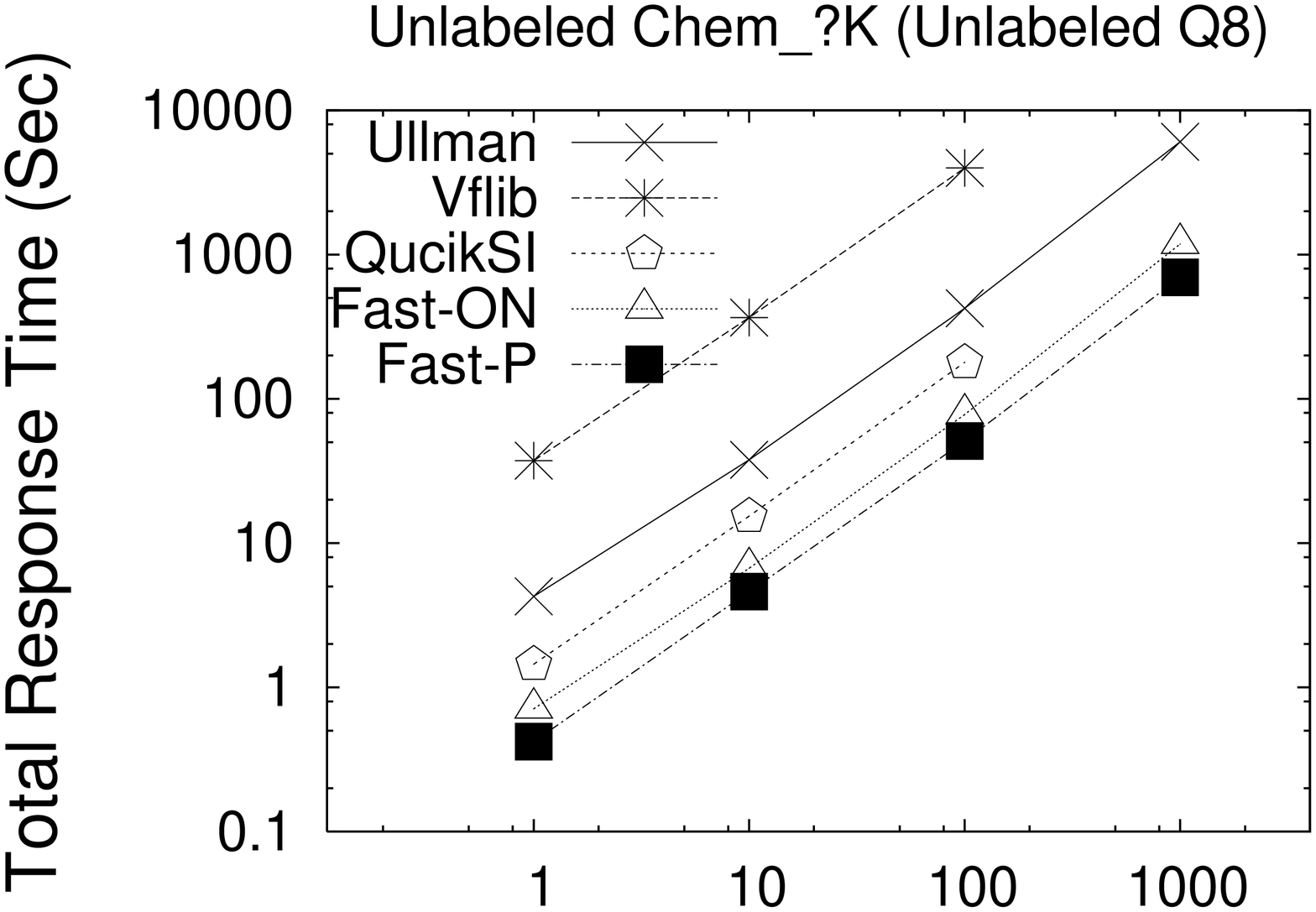,width=1.9in,height=4in,angle=-90}
\caption{Scalability on Dataset Size (\# Graphs in K)}
\label{fig:unlabeled_chemscalability}
\end{figure}
\subsubsection{Scalability} In this experiment, we show the scalability
of Ullman, Vflib, {\tt Fast-ON}, and {\tt Fast-P} on labeled sparse
datasets and the scalability  of Ullman, Vflib, QuickSI, {\tt
Fast-ON}, and {\tt Fast-P} on unlabeled sparse datasets  as follows.
\begin{itemize}
\item {\emph{On Labeled Sparse Datasets}}\\ Figure
\ref{fig:chemscalability} shows the scalability  of Ullman, Vflib,
{\tt Fast-ON}, and {\tt Fast-P} with respect to the number of graphs
using the labeled sparse dataset Chem\_1M and the labeled query set
$Q8$. The figure shows that the four algorithms scale linearly.
However,  {\tt Fast-ON} outperforms Ullman by factor three, and
Vflib by more than one order of magnitude. Moreover,  Vflib is the
worst one and it is not shown for 1000K graphs, since it failed to
run on large datasets. The figure also shows that {\tt Fast-P} has
the best performance, it outperforms Ullman, Vflib, {\tt Fast-ON} by
up to one order of magnitude, more than two order of magnitude, and
up to two factors, respectively.
\item{\emph{On Unlabeled Sparse Datasets}}\\ In this subsection, we
used the Chem\_1M dataset and the query set Q8 after removing the
edge labels and we denoted them as Unlabeled Chem\_1M and Unlabeled
Q8. Figure \ref{fig:unlabeled_chemscalability} shows the scalability
of Ullman, Vflib, QuickSI, {\tt Fast-ON}, and {\tt Fast-P} with
respect to the number of graphs using the sparse dataset Unlabeled
Chem\_1M and the query set Unlabeled $Q8$.  The figure shows that
the five algorithms scale linearly.  However,  {\tt Fast-ON}
outperforms Ullman by factor five, Vflib by more than one order of
magnitude, and QuickSI by up to two factors . Moreover,  Vflib is
the worst one. Note that Vflib and QuickSI are not shown for 1000K
graphs, since they failed to run on large datasets. The figure also
shows that {\tt Fast-P} has the best performance, it outperforms
Ullman, Vflib, QuickSI, and {\tt Fast-ON} by up to one order of
magnitude, up to two order of magnitude, up to four factors, and  up
to two factors, respectively.
\end{itemize}

\section{Conclusion}
\label{sec:Con}
This paper presented two improvements to the Ullmann algorithm, a well-known subgraph isomorphism checker, named {\tt Fast-on} and {\tt Fast-p}. {\tt Fast-on} improves Ullman by reducing its search space using first a refined  vertex matching process and second a new search ordering methodology.  {\tt Fast-p}, on the other hand, is a path-at-a-time matching, leverages structure instead of vertex matching, and uses efficient path ordering methodology to reduce the search space. Experiments show that significant improvements, up to four orders of magnitude, are achieved.
\bibliographystyle{ACM-Reference-Format-Journals}
\bibliography{fast-on-path}



\medskip

\end{document}